\documentclass[showpacs,amssymb,10pt,reprint,aps,prd,longbibliography,nofootinbib,superscriptaddress,floatfix,nobibnotes]{revtex4-1}

\usepackage{graphicx,epsfig,amssymb} 
\usepackage{amsmath,amsfonts, times}
\usepackage{bm} 

\usepackage{epstopdf}

\usepackage{booktabs}

\usepackage{xcolor}

\usepackage[linktocpage,colorlinks]{hyperref}
\usepackage[caption=false]{subfig}
\usepackage{natbib}
\usepackage{subfig}
\usepackage[utf8x]{inputenc}
\usepackage{enumerate}
\usepackage[normalem]{ulem}

\begin{document}

\title{\large Spherical Einstein-Friedberg-Lee-Sirlin boson stars: Self-interacting solutions and their astrophysical appearance}

\author{Pedro L. Brito de Sá}
\date{\today}
\email{pedro.sa@icen.ufpa.br}
\affiliation{Programa de P\'os-Gradua\c{c}\~{a}o em F\'{\i}sica, Universidade 
	Federal do Par\'a, 66075-110, Bel\'em, Par\'a, Brazil.}

\author{Haroldo C. D. Lima}
\email{haroldo.lima@ufma.br}
\affiliation{Programa de P\'os-Gradua\c{c}\~{a}o em F\'{\i}sica, Universidade Federal do Maranh{\~a}o, 65080-805, S{\~a}o Lu{\'i}s, Maranh{\~a}o, Brazil.}
	
\author{Carlos A. R. Herdeiro}
\email{herdeiro@ua.pt}
\affiliation{Departamento de Matem\'atica da Universidade de Aveiro and Centre for Research and Development  in Mathematics and Applications (CIDMA), Campus de Santiago, 3810-183 Aveiro, Portugal.}
\affiliation{Programa de P\'os-Gradua\c{c}\~{a}o em F\'{\i}sica, Universidade 
	Federal do Par\'a, 66075-110, Bel\'em, Par\'a, Brazil.}

\author{Lu\'is C. B. Crispino}
\email{crispino@ufpa.br}
\affiliation{Programa de P\'os-Gradua\c{c}\~{a}o em F\'{\i}sica, Universidade 
	Federal do Par\'a, 66075-110, Bel\'em, Par\'a, Brazil.}

\begin{abstract}

We investigate boson stars within the framework of the self-interacting Einstein–Friedberg–Lee–Sirlin (E–FLS) model, constituted by a complex scalar field with a quartic self-interaction and a real scalar field. Our analysis explores the family of static solutions across a broad range of parameters, including the self-interaction of the complex scalar field. We obtain that positive self-interaction terms increase the maximum mass and compactness of E–FLS stars, allowing them to reach masses comparable to the Chandrasekhar limit without the need of ultralight bosonic masses. Moreover, in the limit where the real scalar field becomes massless, the solutions present larger effective radii and allow a broader range of stable solutions. Astrophysical images, generated via backward ray-tracing, show that these compact, self-interacting E–FLS stars produce strong gravitational lensing, yielding shadows that could visually mimic black holes, thus providing potential observational signatures detectable in ongoing electromagnetic surveys.

\end{abstract}

\maketitle


\section{Introduction}
\label{sec:int}
The current state of observational data within the framework of General Relativity (GR) has seen notable advancements in recent years. The LIGO collaboration has successfully detected gravitational waves (GWs) emitted by the merger of compact objects~\cite{ligo1}. The detection of GWs by the LIGO collaboration initiated a new phase of astronomical observation through a different radiation channel~\cite{ligo2}. Approximately 300 mergers of compact binary systems (interpreted as black hole-black hole, black hole-neutron star, and neutron star-neutron star) have been observed via GWs in the four observation runs of the (now) LIGO-Virgo-KAGRA collaboration.  Furthermore, the Event Horizon Telescope (EHT) collaboration obtained the first images of the central regions of galaxies M87, released in 2019~\cite{eht} and Milky Way, released in 2022~\cite{eht_mw}, using Very Long Baseline Interferometry (VLBI) technique. The images show an inner dark region consistent with the shadow of a supermassive compact object.

These observations strongly corroborate the existence of compact objects, such as black holes. But they also open the possibility for alternative horizonless compact objects exhibiting characteristics closely resembling black holes, which are called black hole mimickers~\cite{bh_mimickers}. Future observatories like the Laser Interferometer Space Antenna (LISA) for low-frequency GWs, the next-generation Event Horizon Telescope (ng-EHT), and the Black Hole Explorer (BHEX) mission are expected to provide data with refined details and increased precision~\cite{bhex}. This data from both gravitational and electromagnetic channels will be crucial for better constraining models that are alternatives to the black hole paradigm~\cite{bh_mimickers2, bh_mimickers3, Mielke2000}.

A prominent class of black hole mimickers are the so-called boson stars (BSs)~\cite{bs_mimickers, Macedo2013, FLS, Kaup}. These stars are composed of bosonic particles, modeled by scalar fields. The BSs that do not present self-interaction, often referred as mini-boson stars (MBSs), are predicted to have maximum masses significantly below the Chandrasekhar limit, given by $\sim 1.4  M_{\odot}$ ~\cite{Macedo2013, lie_pale, Schunck2003}, when the fundamental boson constituent is of a Standard Model-range mass.\footnote{Considering hypothetical ultralight bosons, on the other hand, makes mini-boson stars much more massive.} Adding a potential term, on the other hand, which characterizes self-interacting boson stars (SBSs) ~\cite{colpi86}, leads to maximum masses comparable to (or even exceeding) the Chandrasekhar mass, for Standard Model mass range scalar bosonic particles, being potentially relevant as astrophysical objects ~\cite{lie_pale, Schunck2003, colpi86}. 

The study of BSs is (in part) motivated by their potential to describe dark matter (DM), estimated to constitute approximately 26$\%$ of the Universe's total composition ~\cite{lie_pale}. It is theorized that DM can form self-gravitating configurations, a concept which BSs naturally embody. A key question, however, is how the fundamental boson gets its mass. A natural mechanism is a Higgs-type mechanism. This is naturally included 
in the Friedberg-Lee-Sirlin (FLS) model. This model, originally introduced as an example of a scalar renormalizable theory, involves a complex scalar field coupled with a real scalar field, where the latter acquires a non-zero vacuum expectation value through a symmetry-breaking mechanism~\cite{FLS}. The interaction between these fields, specifically through a coupling term, is responsible for generating their effective masses ~\cite{FLS}. When the FLS model is minimally coupled to Einstein’s gravity, it is called Einstein-Friedberg-Lee-Sirlin (E-FLS) theory ~\cite{Rotating_FLS, U1_FLS}. Previous investigations were dedicated to the study of E-FLS theory without a self-interaction term in the complex scalar field sector~\cite{Rotating_FLS, Sa2024}. In the present work, we explore the E-FLS theory with a quartic self-interaction term in the complex scalar field, crucial for BSs to achieve masses comparable to the Chandrasekhar limit~\cite{Macedo2013},  without invoking ultralight masses. Indeed, as we shall show, these self-interacting  E-FLS stars can attain maximum masses that are comparable to, or even exceed, the Chandrasekhar mass, thereby establishing them as potentially significant astrophysical objects. Furthermore, the existence of self-interaction in BSs can also work as a stabilization mechanism~\cite{Sanchis-Gual2022}. 




This paper is organized as follows. In Sec.~\ref{sec:simodel}, we explore the self-interacting E-FLS theory, and define the setup used to obtain the solutions. In Sec.~\ref{sec:numerical_resul}, we present the numerical tools utilized to generate the solutions and provide a complete analysis of the solution space in the static case, exploring the full range of parameters. In Sec.~\ref{sec:geo_ana}, the timelike and null geodesics around these stars are analyzed, including circular orbits. Finally, in Sec.~\ref{sec:astro_images}, we obtain the astrophysical images of these stars, when surrounded by an accretion disks, using the backwards ray-tracing method. The concluding discussion is provided in Sec.~\ref{sec:conclusions}. Throughout the paper, we employ geometrized units with $c=\hbar=1$ and use a metric signature of~$+2$

\section{The self-interacting E-FLS theory}
\label{sec:simodel}
We investigate soliton-like solutions in $(3+1)$-dimensions with two scalar fields: a complex field and a real field. These fields are minimally coupled to gravity, and their interaction is governed by the FLS model~\cite{FLS}. In the case considered here, the complex scalar field possesses a self-interaction term. The self-interacting E-FLS model is defined by the following action:
\begin{align}
\label{action}S=\int\left(\frac{R}{2\kappa^2}-\mathcal{L}_{m}\right)\sqrt{-g}\,d^4x, \qquad \kappa^2\equiv 8\pi G,
\end{align}
in which the quantities $R$ and $g$ are, respectively, the Ricci scalar and the determinant of the metric tensor. The matter Lagrangian is given by:  
\begin{align}
&\mathcal{L}_{m}=\frac{1}{2}\nabla_\mu\psi\, \nabla^\mu\psi+\nabla_\mu\Phi\,\nabla^\mu \Phi^*+U(\psi, \Phi),\label{lm}\\
&U(\psi, \Phi)=m^2\psi^2\left|\Phi\right|^2+ \frac{\lambda}{2}\left|\Phi\right|^4+\mu^2\left(\psi^2-v^2\right)^2, \label{pot}
\end{align}
where $\psi$ and $\Phi$ are self interacting real and complex scalar fields, respectively. The term $m^2\psi^2|\Phi|^2$ couples the two fields, while the $\lambda$ term characterizes a self-interaction contribution for the complex scalar field. This self-interacting behavior can be negative or positive. The constant $v$ corresponds to the vacuum expectation value of the real scalar field. The potential $U(\psi, \Phi)$, with its coupling term, gives rise to the effective masses for the real and complex scalar fields. By expanding $U(\psi, \Phi)$ around $\psi = v$ and $\Phi = 0$, the masses of the scalar fields can be determined: $m_\psi = \sqrt{8} \mu v$ for the real field and $m_\phi = m v$ for the complex field. The masses of the scalar fields are independent of $\lambda$, since the quartic self-interaction term provides only subleading contributions near $\Phi = 0$. In the limit where $\mu \to \infty$, the real scalar field becomes extremely massive and effectively separates from the complex field, with $\psi$ approaching $v$, reducing the E-FLS theory to the Einstein-Klein-Gordon (EKG) model with self-interaction term, as studied by Colpi, Shapiro and Wasserman in~\cite{colpi86}. On the other hand, when $\mu \to 0$, the real scalar field becomes massless, and the resulting E-FLS star solutions deviate significantly from those of Ref.~\cite{colpi86}.


Varying the action~\eqref{action} with respect to the metric tensor and the scalar fields, we obtain the self-interacting E-FLS field equations. The Einstein's field equations are given by:
\begin{align}
\label{EFLS}R_{\mu\nu}-\frac{1}{2}R\,g_{\mu\nu}=\kappa^2 T_{\mu\nu},
\end{align}
with the stress-energy tensor of the scalar fields written as:
\begin{align}
T_{\mu\nu}=\nabla_\mu\Phi\nabla_\nu\Phi^*+\nabla_\mu\Phi^*\nabla_\nu\Phi+\nabla_\mu\psi\nabla_\nu\psi-\mathcal{L}_m\,g_{\mu\nu}.
\end{align}
The variation of the action with respect to the complex scalar field $\Phi$ and the real scalar field $\psi$ gives, respectively: 
\begin{align}
\label{eom_cfield}&\square\Phi=m^{2}\psi^2\Phi + \frac{\lambda}{2} \left|\Phi\right|^{2} \Phi,\\
\label{eom_rfield}&\square\psi=2\psi\left(m^2\left|\Phi\right|^2+2\mu^2\psi^2-2\,v^2\,\mu^2\right),
\end{align}
with $\square\equiv \nabla^\mu\,\nabla_\mu$ being the d'Alembertian operator. The constants $v$ and $m$ appearing in Eqs.~\eqref{EFLS}–\eqref{eom_rfield} can be removed from the equations by introducing the rescaling below::
\begin{align}
\nonumber &r=\frac{\tilde{r}}{mv}, \quad \psi=v\tilde{\psi},\quad \Phi=v\tilde{\Phi},\quad \omega=mv\tilde{\omega},\\ \label{rescal} &\kappa=\frac{\tilde{\kappa}}{v},\quad \mu=m\tilde{\mu},\quad \lambda=m\tilde{\lambda}, 
\end{align}
making it equivalent to set $v=m=1$ in \eqref{eom_cfield}–\eqref{eom_rfield}. In what follows, all equations are expressed in their rescaled form, and we leave out the tildes to keep it simple.

The action \eqref{action} is invariant under the global $U(1)$ transformation $\Phi \to e^{i\alpha}\Phi$, which implies a continuous symmetry. According to Noether’s theorem, this continuous symmetry leads to a conserved current given by:
\begin{align}
j_{\mu}= i\left(\Phi\,\nabla_{\mu}\Phi^{*}-\Phi^{*}\,\nabla_{\mu}\Phi \right), \label{J}
\end{align}  
satisfying $\nabla_{\mu}j^{\mu}=0$. Integrating the component $j^0$ over a spacelike hypersurface $\Sigma$ gives the Noether charge:
\begin{align}
N\equiv \frac{1}{4\pi}\int_{\Sigma} j^{0}\sqrt{-g}d^3x. \label{nnum}
\end{align} 
This quantity can be interpreted as the total number of scalar particles that compose the system.

In this work, we investigate the E-FLS spherically symmetric star solutions with self-interaction. To find the BS solutions, we first need to choose a spherically symmetric Ansantz, which can be given by the following line element in Schwarzschild-like coordinates:
\begin{align}
\label{metric}ds^2=-e^{\Gamma(r)}dt^2+e^{\Lambda(r)}dr^2+r^2\left(d\theta^2+\sin^2\theta\,d\varphi^2\right),
\end{align}
with $\Gamma(r)$ and $\Lambda(r)$ metric functions with only radial dependence, since the system is static and spherically symmetric. 
We also need to choose a spherically symmetric Ansatzë for the real and complex scalar fields: 
\begin{align}
\label{r_field}&\psi=\psi(r),\\
\label{c_field}&\Phi(t, r)=\phi(r)e^{i\omega\,t},
\end{align}
respectively. The functions $\phi(r)$ and $\psi(r)$ are purely real and possess dependence only on the coordinate $r$. $\omega$ is a real and positive parameter associated with the oscillating frequency of the complex scalar field. Substitution of Eqs. \eqref{metric}–\eqref{c_field} into the E–FLS field equations \eqref{EFLS}–\eqref{eom_rfield} yields the following system of coupled ordinary differential equations (ODEs):
\begin{align}
\nonumber &\frac{e^{-\Lambda}\,\Lambda'}{r}+\frac{\left(1-e^{-\Lambda}\right)}{r^2}=\kappa^2\left[\omega^2\,e^{-\Gamma}\phi^2 \right. \\
\label{ode_gtt} & \left. +e^{-\Lambda}\phi'^2+\frac{e^{-\Lambda}\,\psi'^2}{2}+\psi^2\phi^2+\frac{\lambda}{2}\phi^{4}+\mu^2\left(\psi^2-1\right)^2 \right],\\
\nonumber &\frac{e^{-\Lambda}\,\Gamma'}{r}-\frac{\left(1-e^{-\Lambda}\right)}{r^2}=\kappa^2\left[\omega^2\,e^{-\Gamma}\phi^2 \right.\\
&\left.+ e^{-\Lambda}\phi'^{2}+\frac{e^{-\Lambda}\,\psi'^2}{2}-\psi^2\phi^{2} - \frac{\lambda}{2}\phi^{4}-\mu^2\left(\psi^2-1\right)^2 \right],\\
\label{scalar_field} &\phi''+\left(\frac{2}{r}+\frac{\Gamma'-\Lambda'}{2}\right)\phi'+e^\Lambda\left(\omega^2 e^{-\Gamma} - \lambda\phi^{2}-\psi^2\right)\phi=0,\\
\nonumber &\psi''+\left(\frac{2}{r}+\frac{\Gamma'-\Lambda'}{2}\right)\psi'-4\mu^2e^{\Lambda}\psi^3\\
\label{psi} &+2\,\left(2\,\mu^2-\phi^2 \right)e^{\Lambda}\psi=0.
\end{align}
In the set of Eqs. \eqref{ode_gtt} to \eqref{psi} the prime indicates differentiation with respect to the radial coordinate $r$.
To obtain the BS solutions in the self-interacting E–FLS framework, we must impose the appropriate boundary conditions to the set of Eqs.~\eqref{ode_gtt} to \eqref{psi}. We aim to obtain configurations that remain regular at the center $(r=0)$ and approach flat spacetime as $r \to \infty$. In this context, the boundary conditions to ensure regularity at the origin are given by:
\begin{align}
\label{bc1}\Lambda'(r=0)=0, \quad \phi'(r=0)=0, \quad \psi'(r=0)=0.
\end{align}
Additionally, the requirement of asymptotic flatness results in the following boundary conditions:
\begin{align}
\label{bc2}\Gamma(r=\infty)=0, \quad \phi(r=\infty)=0, \quad \psi(r=\infty)=1.
\end{align}

Given a fixed value of the scalar field at the origin, $\phi(0) \equiv \phi_0$, the system of equations \eqref{ode_gtt}–\eqref{psi}, and the boundary conditions \eqref{bc1}–\eqref{bc2}, represents  an eigenvalue problem for the frequency $\omega$. Each value of the complex scalar field at the origin $\phi_0$ is associated with an infinite set of eigenvalues ${\omega_n}$. The integer $n$ denotes the number of nodes in the complex scalar field. Excited states, corresponding to $n \geq 1$, are generally unstable for mini-BSs~\cite{Gleiser1988, Balakrishna1998}, but can be stabilized by the quartic self-interactions we are considering~\cite{Sanchis-Gual:2021phr,Brito:2023fwr}. The stability of the fundamental mode ($n = 0$) in the FLS theory was analyzed in Ref.~\cite{FLS} for solitons in flat spacetime. 

In this work, we focus on the fundamental solutions with $n = 0$. In the following section, we outline the numerical method used to compute the nodeless solutions and present a selection of numerical results obtained in our investigation.
\section{Results}
\label{sec:numerical_resul}

\subsection{The numerical method}

To numerically solve the two-point boundary value problem outlined in Sec.~\ref{sec:simodel}, we employ the FORTRAN package COLSYS (Collocation for Systems), originally introduced in Ref.~\cite{colsys}, which utilizes a spline collocation method at Gaussian points to solve boundary value problems for ordinary differential equations (ODEs). The numerical solutions are obtained by providing an initial approximation close to the true solution, starting with large values of the parameter $\mu$ and small values of the self-interaction term $\lambda$, given that we can recover the MBS case of the EKG model when $\mu\rightarrow \infty$ and $\lambda = 0$. The MBS initial guess was constructed using a standard shooting method~\cite{Cpp}. We first compute the solutions of E-FLS stars without self-interaction term (given by $\lambda = 0$) and then we utilize such solutions as initial guesses to compute the case where $\lambda \neq 0$. After obtaining the solution for large $\mu$, we gradually decrease $\mu$ until reaching $\mu = 0$, and due to the invariance of the field equations under $\mu \rightarrow -\mu$, it is unnecessary to compute solutions for negative values of $\mu$, enabling a complete exploration of the solution space within the E-FLS theory. In this work, we implemented the numerical method using a grid of $2000$ points and achieved a relative precision on the order of $10^{-12}$.

\subsection{Self-interacting E-FLS stars}
\begin{figure*}[h!]
\centering
\subfloat{
\includegraphics[scale=0.4]{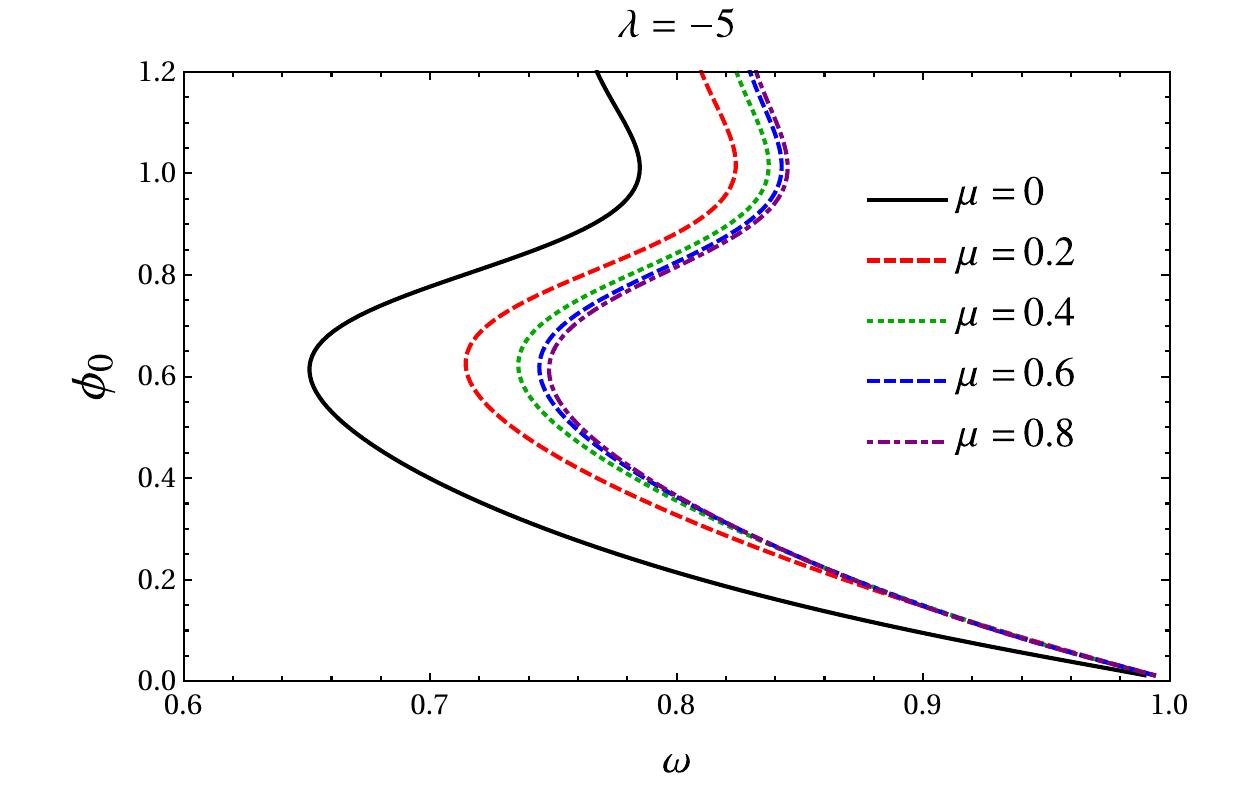}}
\subfloat{
\includegraphics[scale=0.4]{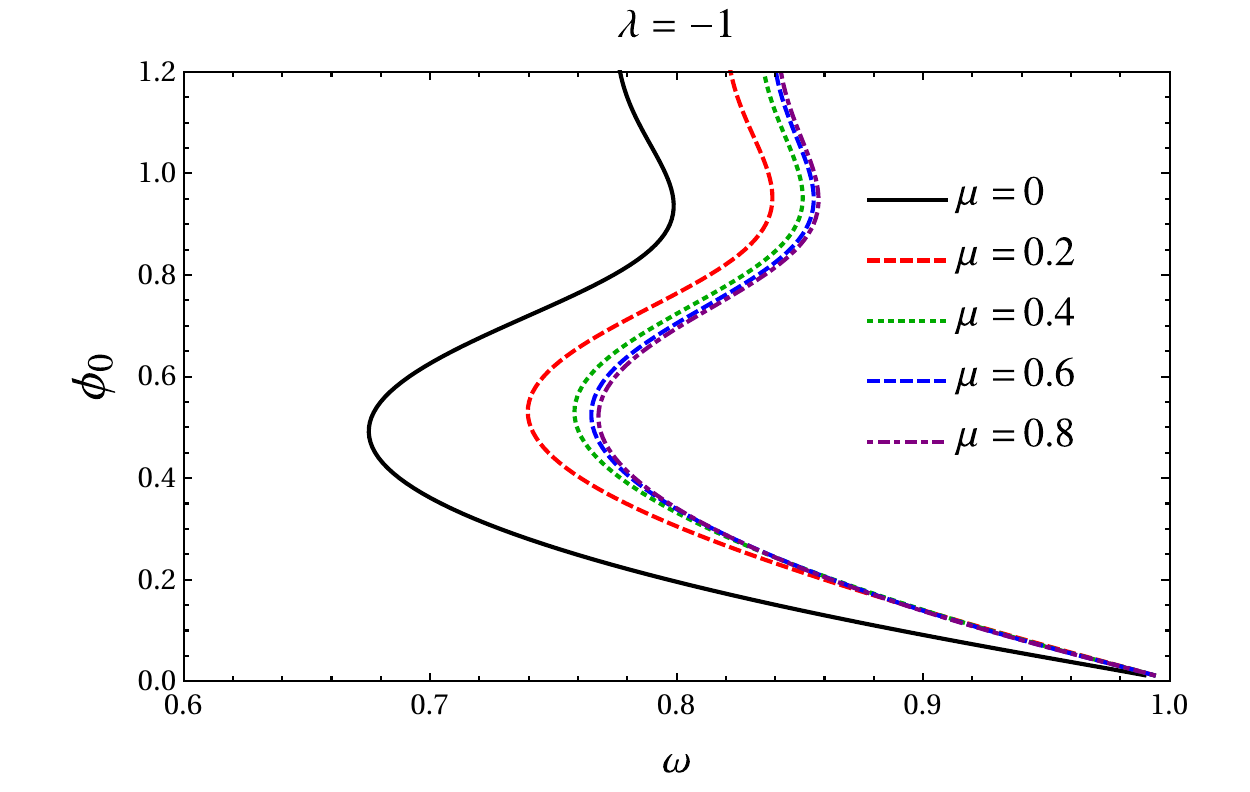}}
\\
\subfloat{
\includegraphics[scale=0.4]{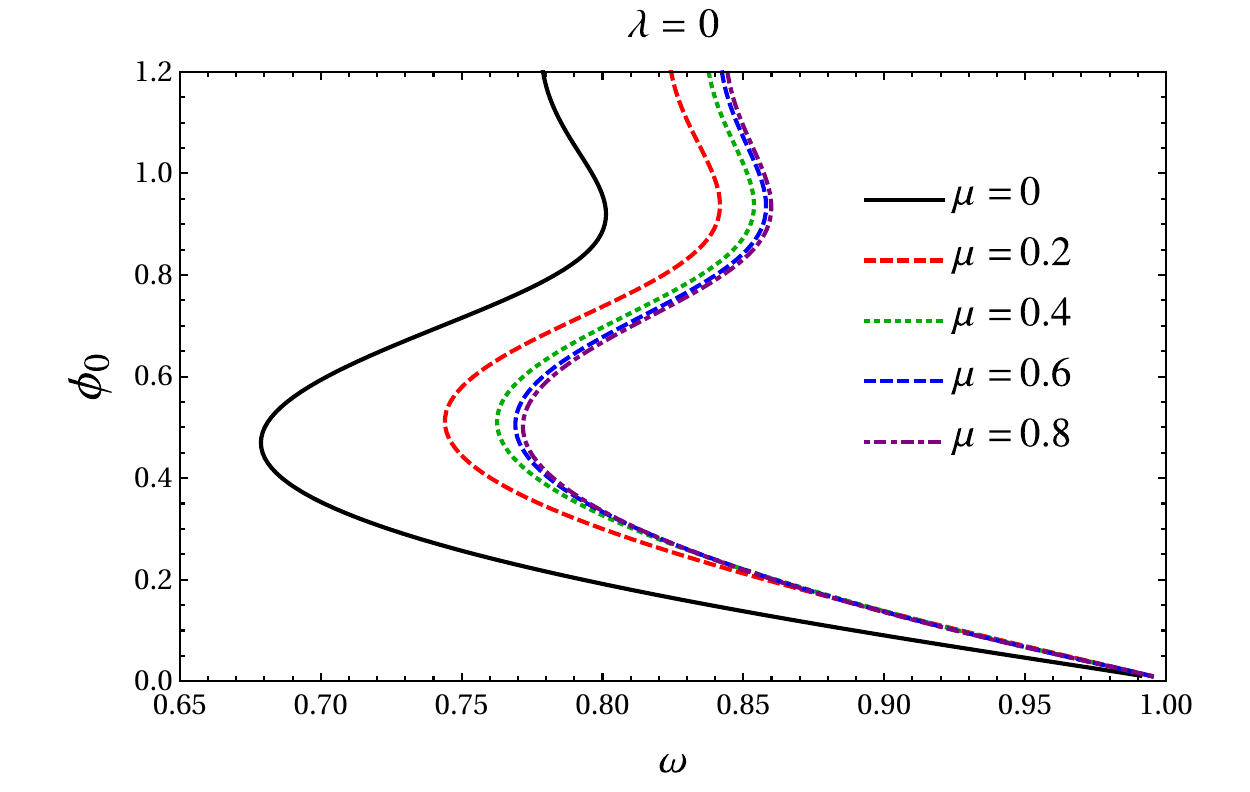}}
\subfloat{
\includegraphics[scale=0.4]{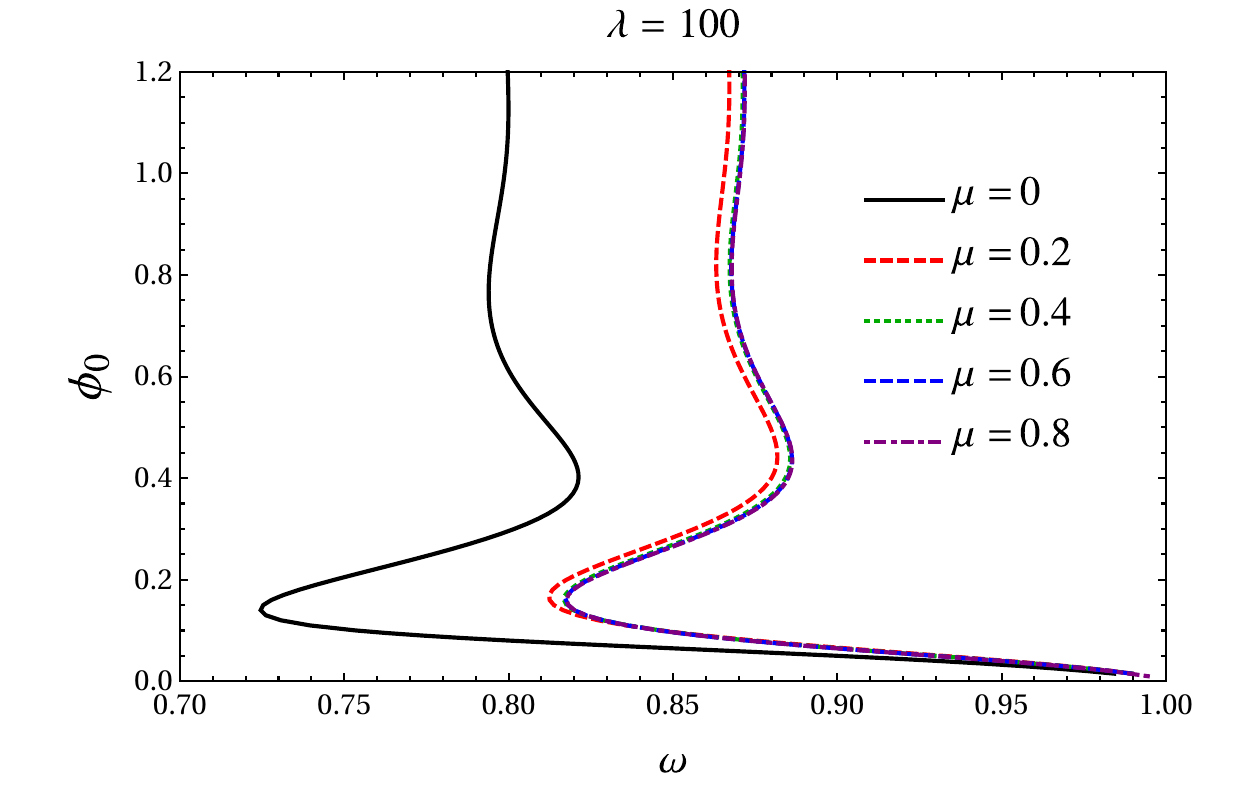}}
\\
\subfloat{
\includegraphics[scale=0.4]{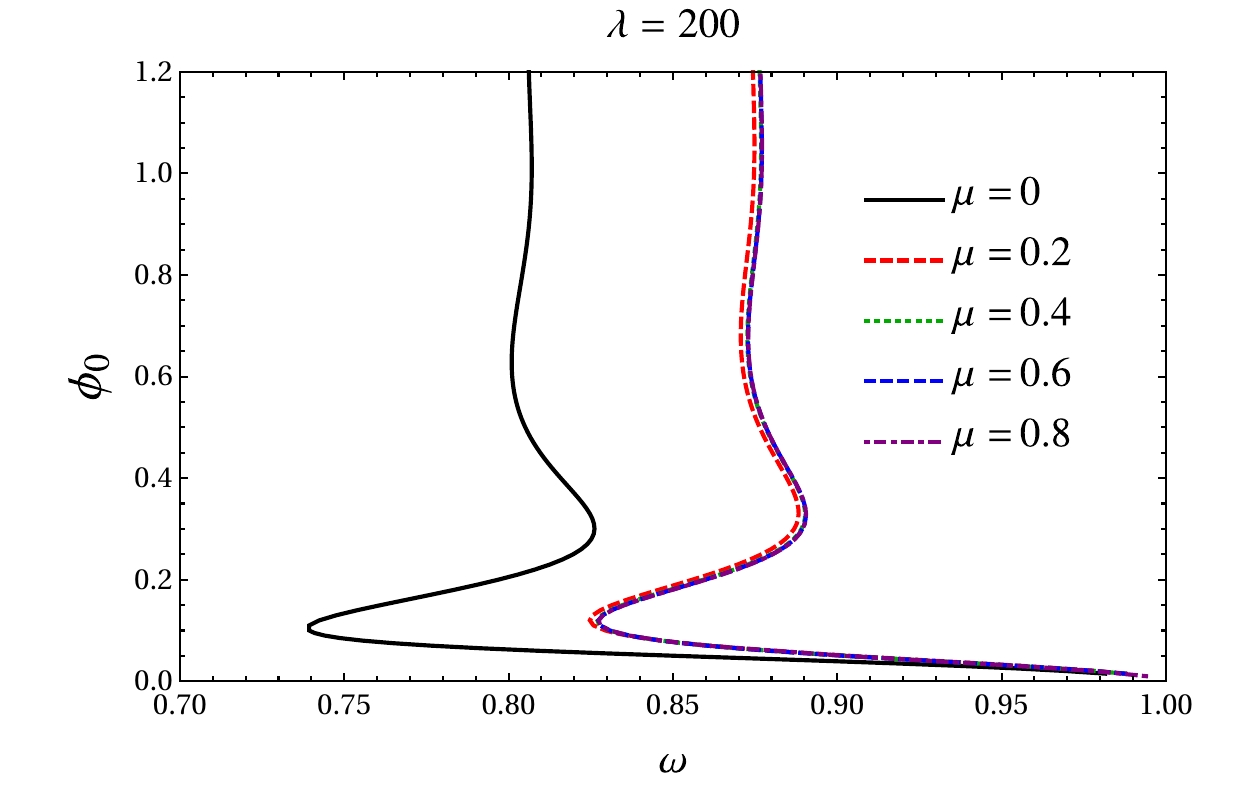}
}
\subfloat{
\includegraphics[scale=0.4]{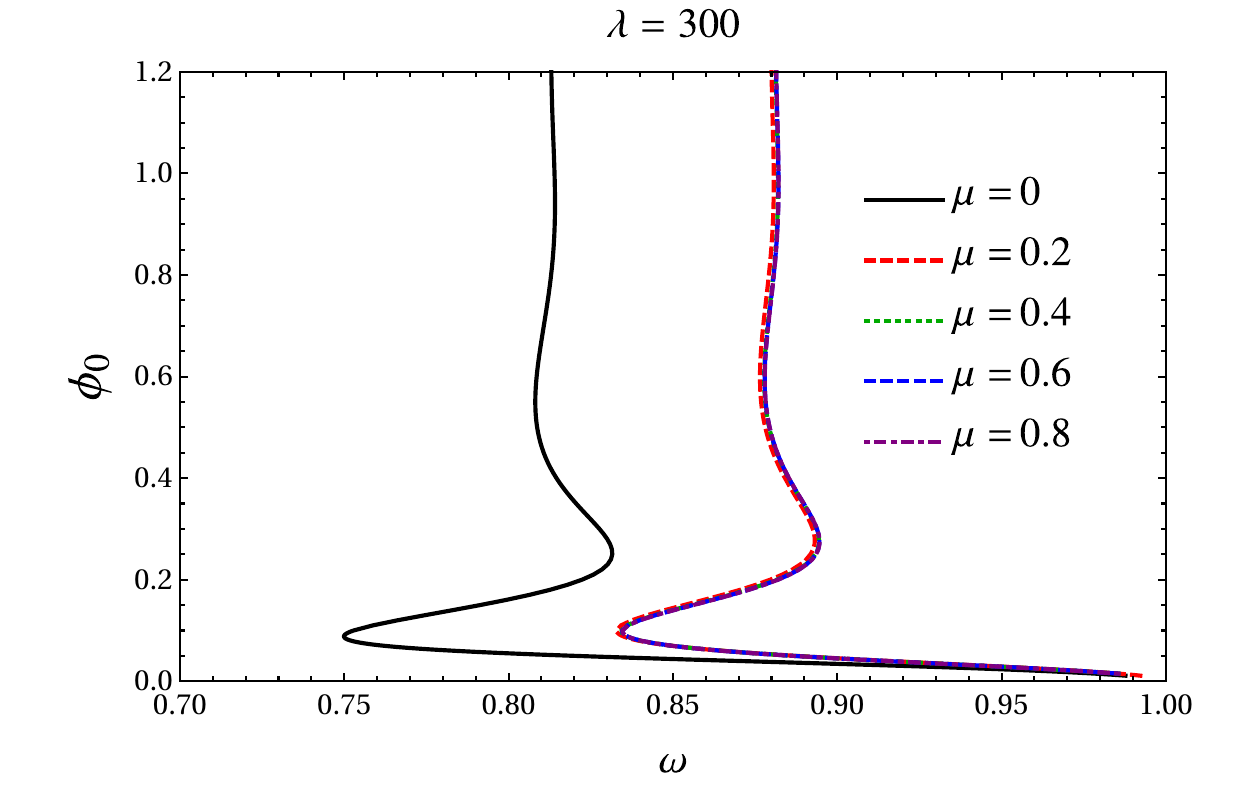}
}
\caption{The complex scalar field at the origin $\phi_{0}$ as a function of the eigenvalues $\omega$ for six different values of the self-interaction term $\lambda$. We also vary the value of the mass parameter $\mu$ to investigate the impact on self-interacting E-FLS stars solutions.}
\label{all_solut}
\end{figure*}

We now present part of our numerical results for BS solutions within the self-interacting E-FLS theory. The numerical solutions were computed for various values of the parameters $\phi_0$, $\mu$, and selected values of $\lambda$ to explore the properties of the solution space. In Fig.~\ref{all_solut}, the full family of E-FLS star solutions is shown, considering variations in $\mu$ and $\lambda$ for different values of the complex scalar field at the origin, $\phi_0$, as a function of the frequency $\omega$. For small values of $\phi_0$, the solutions approach the Newtonian regime, where the E-FLS stars are less compact and the oscillation frequency $\omega$ is close to $1$~\cite{Friedberg1987b}. As $\phi_0$ increases, the E-FLS stars become more compact, and relativistic effects begin to dominate. For instance, pairs of circular photon orbits can emerge in this regime~\cite{Cunha:2017wao,Cunha2017}. 
The E-FLS stars exist within the frequency range $\omega_{\min} \leq \omega \leq 1$, where the upper bound corresponds to the Newtonian limit, and the lower bound $\omega_{\min}$ depends on the specific value of the FLS parameters $\mu$ and $\lambda$. As the self-interaction term $\lambda$ varies, the profiles of the solutions change. For $\lambda < 0$, the set of solutions shift to the left, diminishing the value of $\omega_{min}$ for all selected $\mu$. Notably, for $\lambda > 0$, the family of solutions with $\mu \neq 0$ becomes increasingly similar, while the set of solutions with $\mu = 0$ maintains its distinct characteristics. This non-trivial behavior for the vanishing $\mu$ arises because the interaction with the complex scalar field is governed by $m_\psi = \sqrt{8}\mu\,v$. When $\mu = 0$, the real scalar field becomes massless, causing the interaction to be long-ranged in the limit $\mu \rightarrow 0$~\cite{Loiko2018}.

Since the spacetime analyzed in this work is spherically symmetric, we can use the following definition for the mass function $\mathcal{M}(r)$:
\begin{equation}
\mathcal{M}(r) = \frac{r}{2}\left(1- e^{-\Lambda(r)}\right). \label{mass}
\end{equation}
The ADM mass of the E-FLS stars is computed by taking the limit $M\equiv\mathcal{M}(r \rightarrow \infty)$. 
Given the definition of the Ans{\"a}tze for the metric and the scalar field, we can use the virial identity as a method to verify the consistency of the computed numerical solutions.

\begin{figure*}[h!]
\centering
\subfloat{
\includegraphics[scale=0.4]{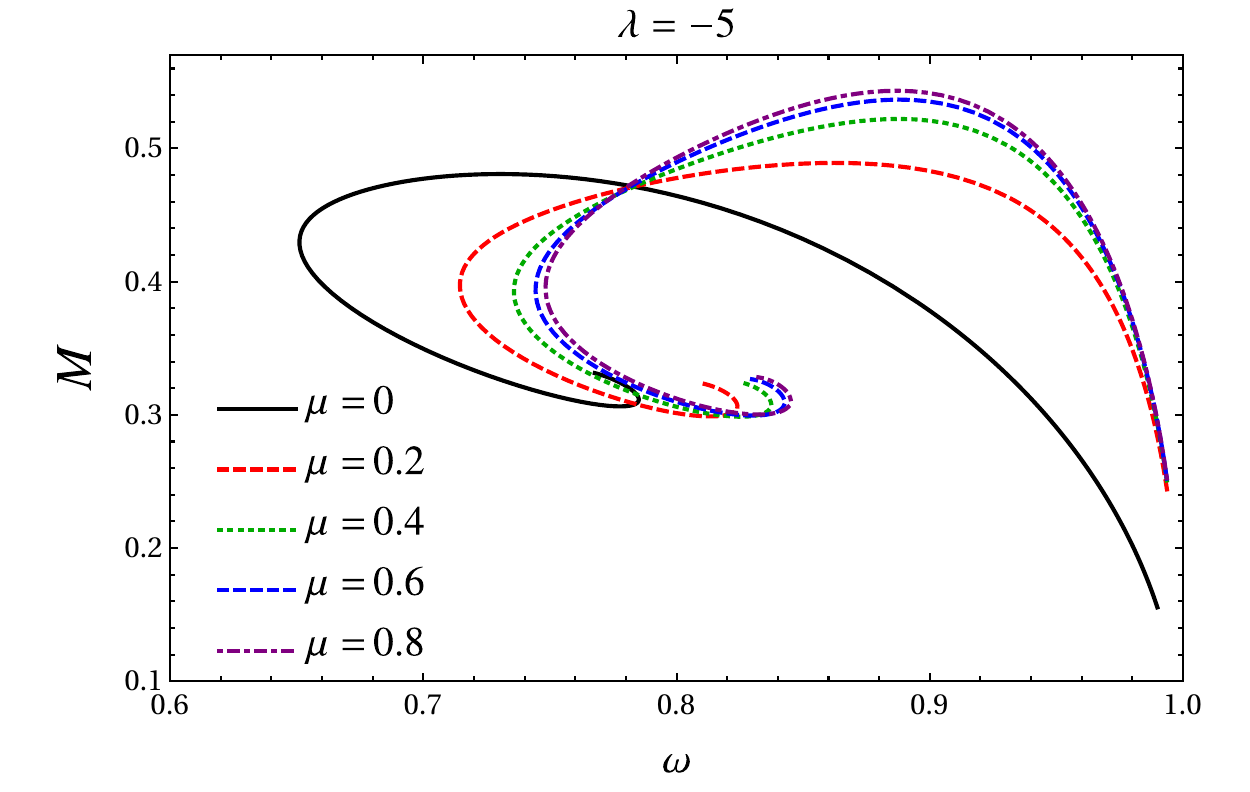}}
\subfloat{
\includegraphics[scale=0.4]{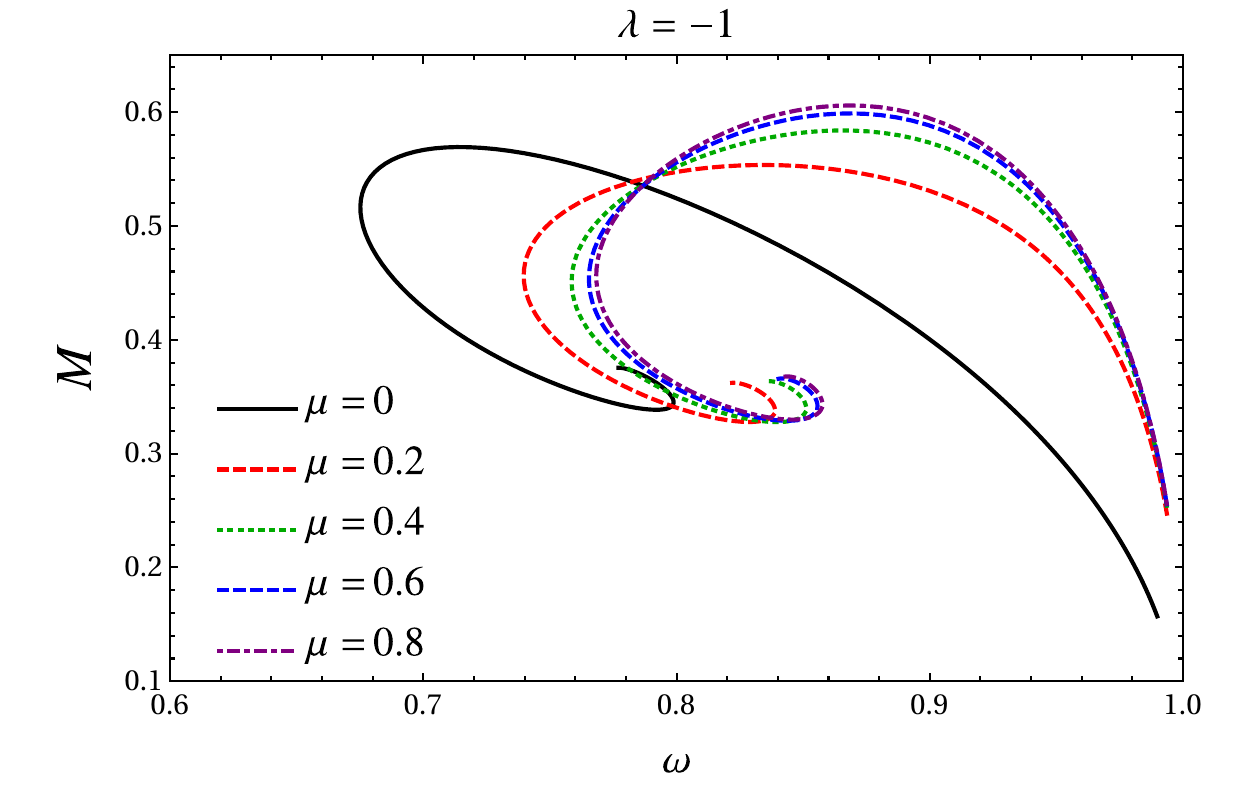}}
\\
\subfloat{
\includegraphics[scale=0.4]{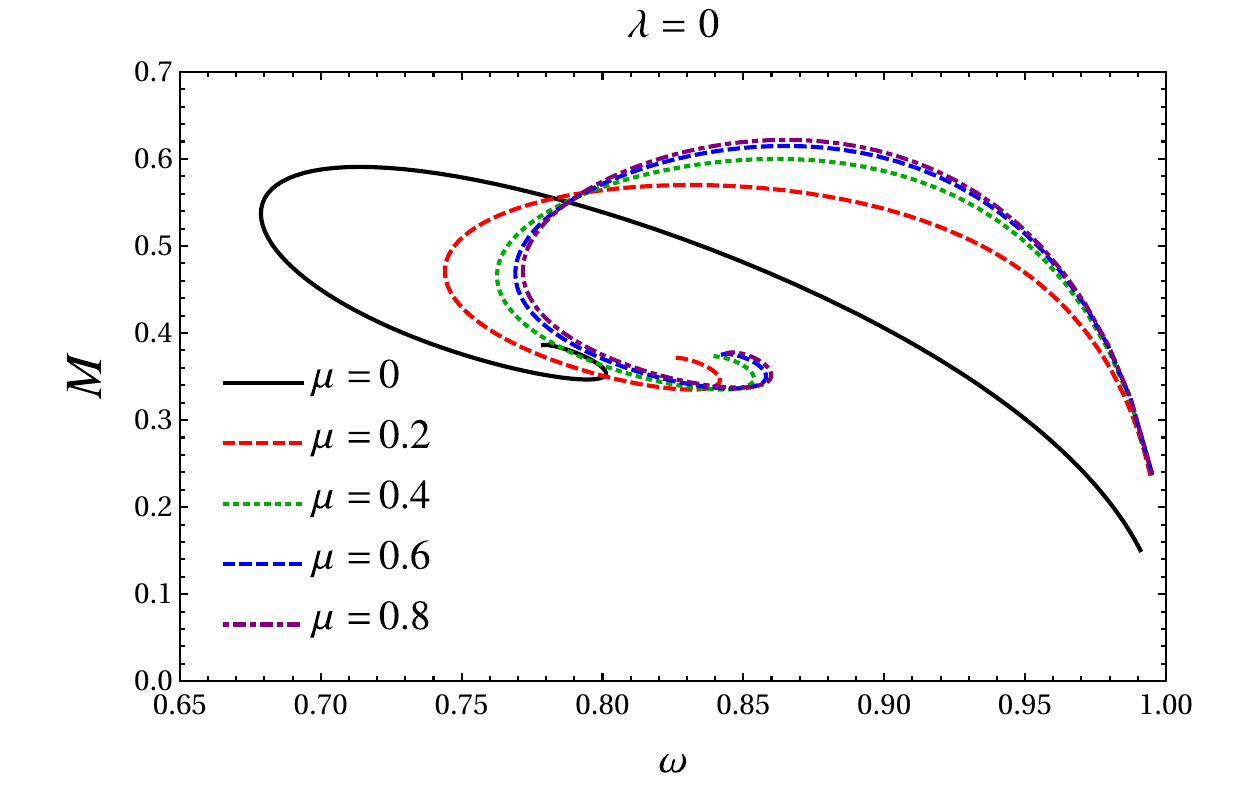}}
\subfloat{
\includegraphics[scale=0.4]{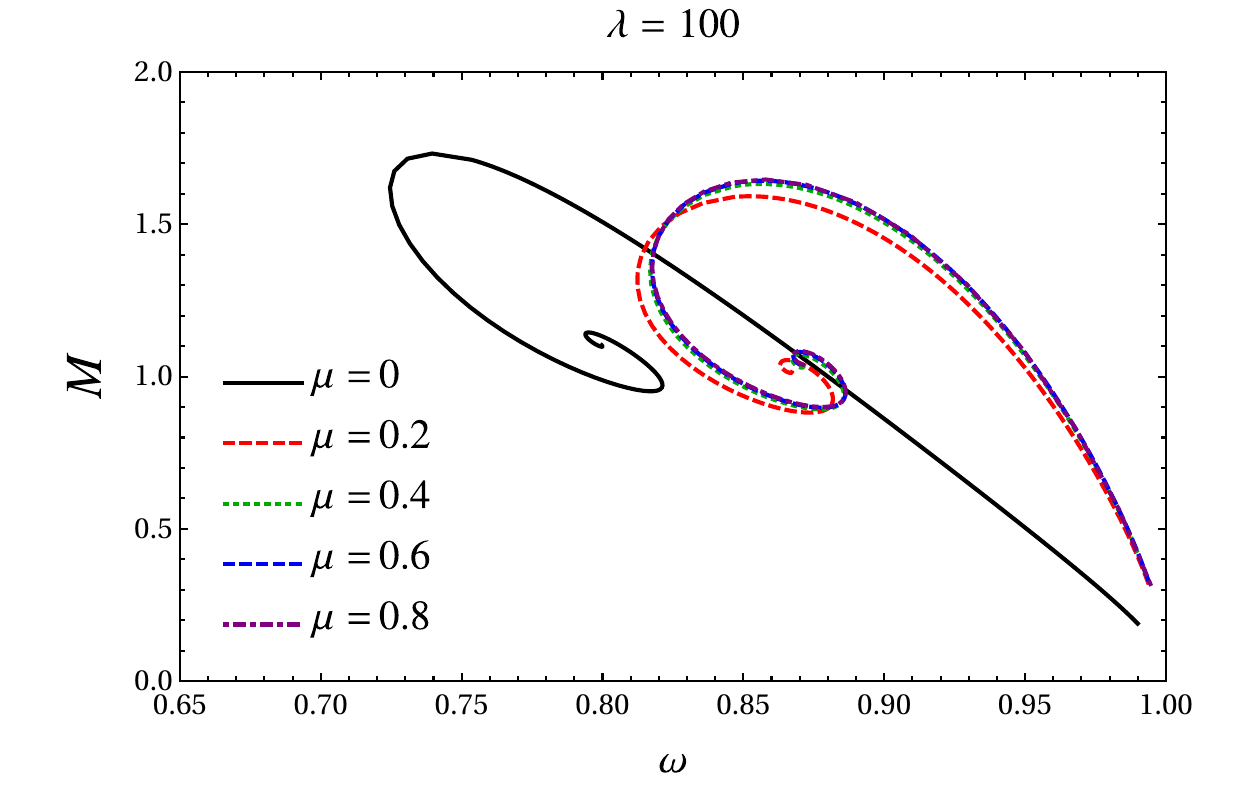}}
\\
\subfloat{
\includegraphics[scale=0.4]{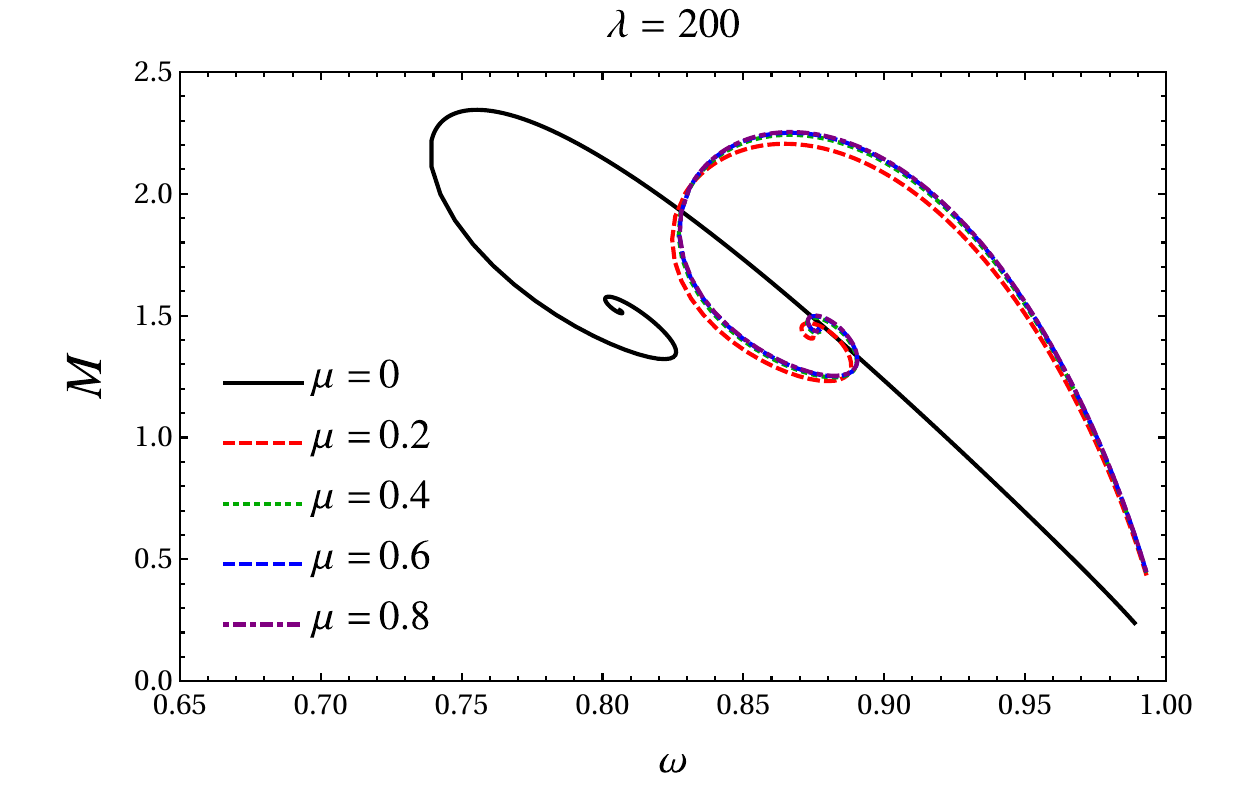}
}
\subfloat{
\includegraphics[scale=0.4]{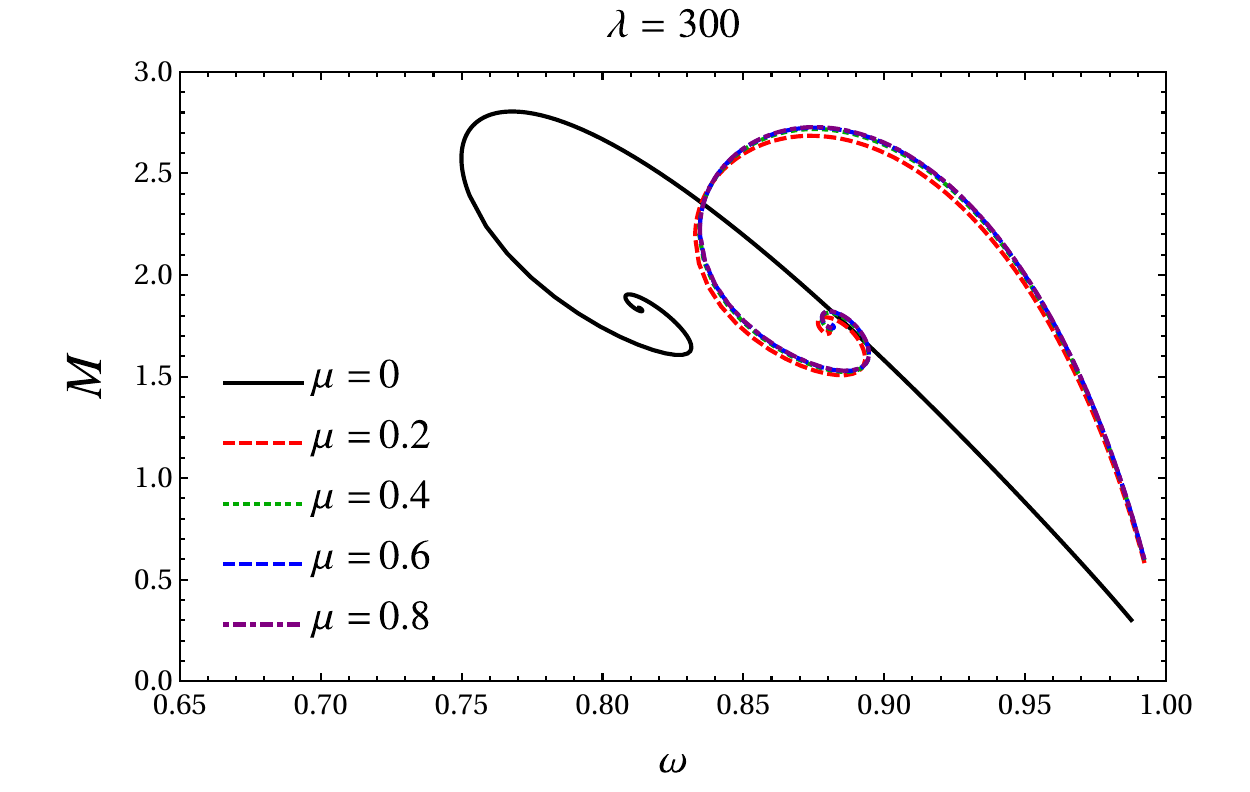}
}

\caption{The total mass $M$ of E-FLS stars, for different values of $\mu$ and $\lambda$, as a function of the normalized frequency ${\omega}$. Each point in these panels represents a self-interacting E-FLS star solution with a given $\phi_0$. The set of solutions with negative values of $\lambda$ are the ones with smaller maximum total mass. Increasing the value of $\lambda$ leads to an increase in the total mass limit of the stars.}
\label{mass_omega}
\end{figure*}

\begin{figure}[h!]
\includegraphics[scale=0.4]{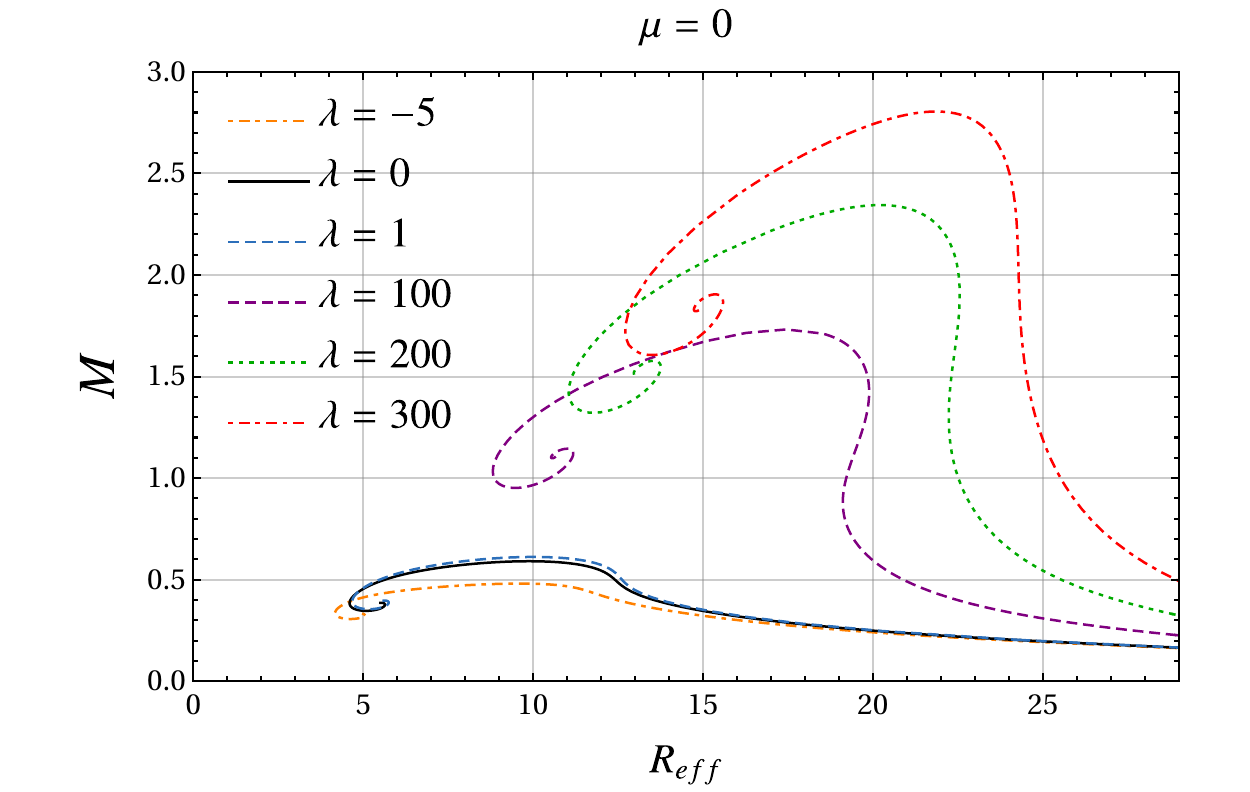}
\caption{The total mass $M$ as a function of the effective radius for five different values of the self-interaction parameter $\lambda$, considering $\mu=0$. Increasing the value of $\lambda$ also increases the total mass of E-FLS stars.}
\label{mxref}
\end{figure}

Fig.~\ref{mass_omega} displays the ADM mass of E-FLS stars as a function of the oscillation frequency $\omega$. Each panel of Fig.~\ref{mass_omega} corresponds to a distinct value of the self-interaction parameter $\lambda$. Negative values of $\lambda$ tend to diminish the maximum total mass of the solutions. As $\lambda$ increases, the maximum mass of the stars also increases, allowing the E-FLS configurations to support more massive solutions. 
The results demonstrate how variations in the parameter $\mu$ influence the total mass profile of the stars. Notably, the case $\mu = 0$ leads to distinct configurations when compared to $\mu \neq 0$. For larger values of $\lambda$, the total mass profiles of solutions with $\mu \neq 0$ become increasingly similar, while the $\mu = 0$ case remains unique. The mass function \eqref{mass} is also useful for defining an effective radius $R_{eff}$ for the E-FLS stars. It is important to note that the scalar field extends to spatial infinity, meaning E-FLS stars do not have a well-defined radius. Therefore, we define the effective radius of the star as the hypersurface containing 99$\%$ of the total ADM mass. In Fig.~\ref{mxref}, we show the profile of the ADM mass as a function of the effective radius $R_{eff}$, considering different values of the parameter $\lambda$ and choosing the parameter $\mu=0$. As we increase the values of the self-interacting term, families of solutions with larger maximum total mass and effective radius emerge.

\begin{figure*}
\centering
\subfloat{
\includegraphics[scale=0.36]{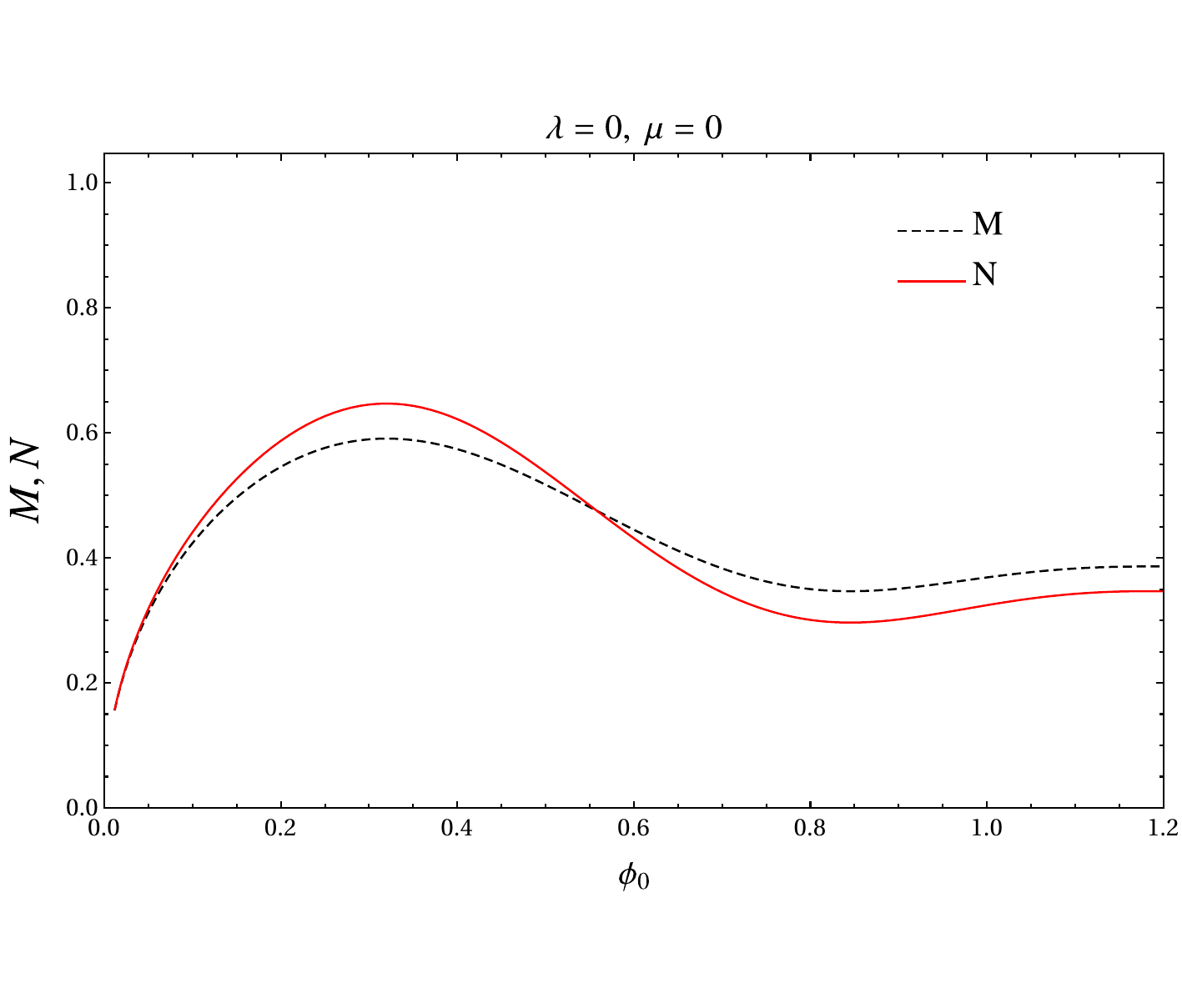}}
\subfloat{
\includegraphics[scale=0.36]{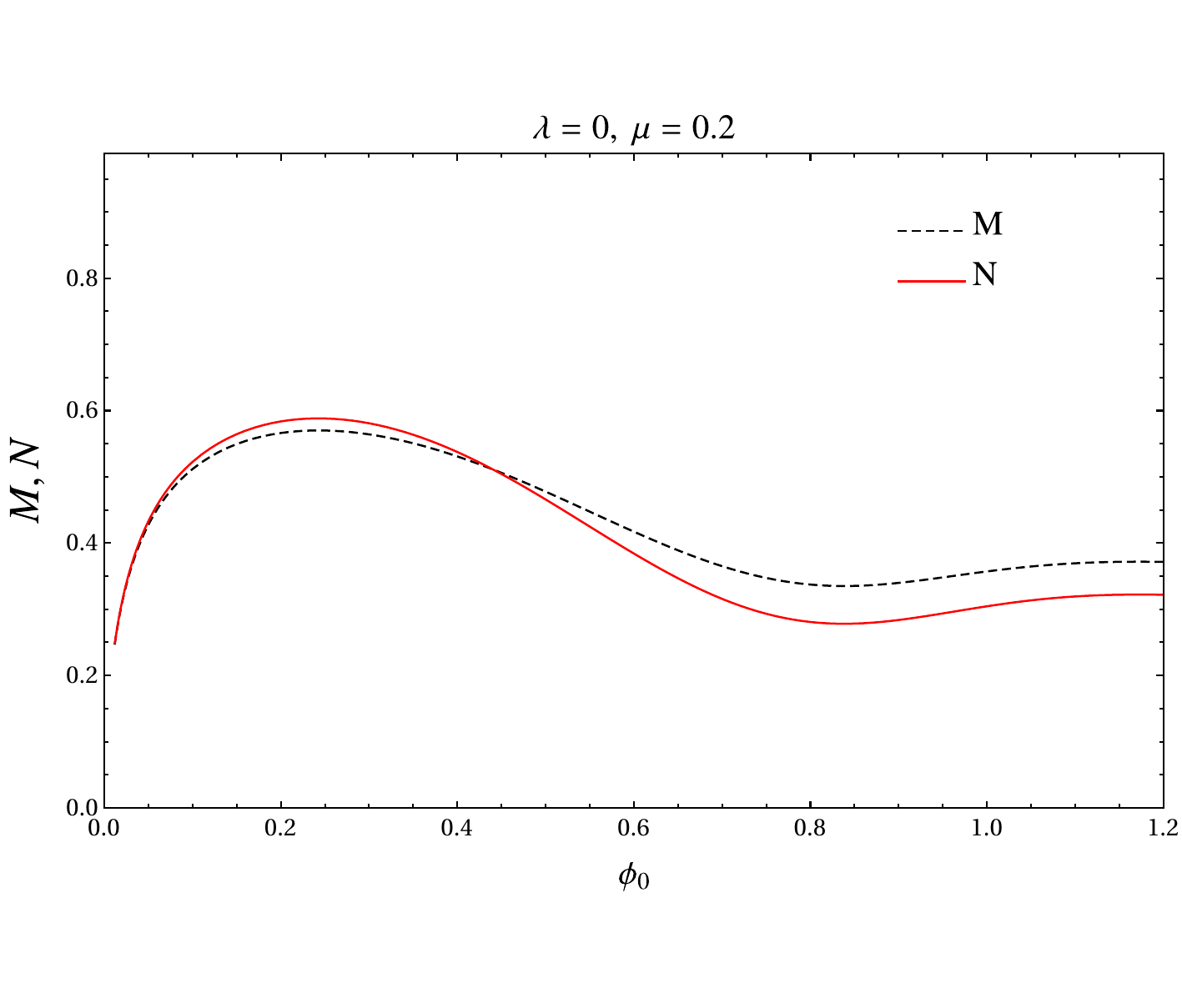}}
\\
\subfloat{
\includegraphics[scale=0.36]{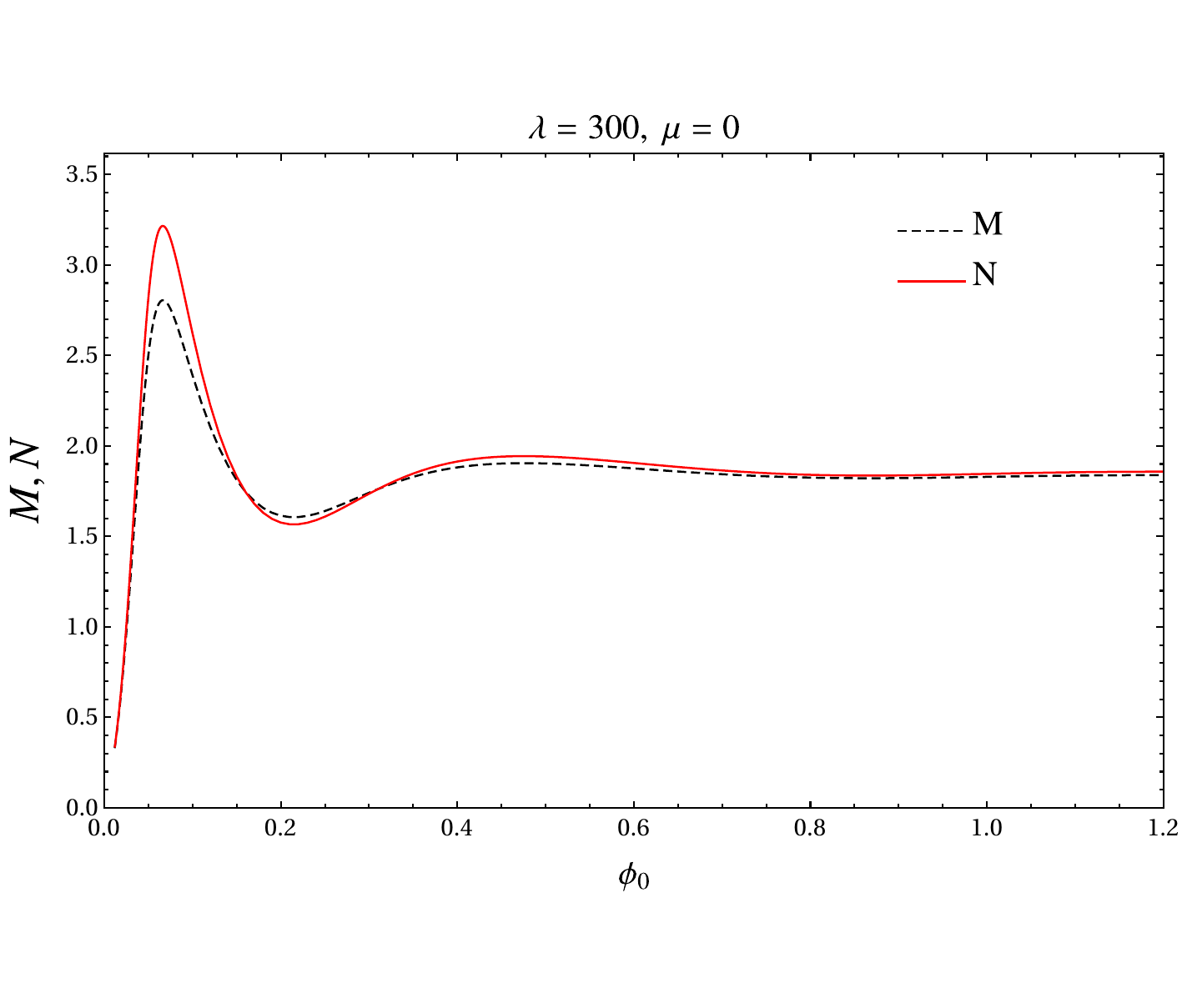}
}
\subfloat{
\includegraphics[scale=0.36]{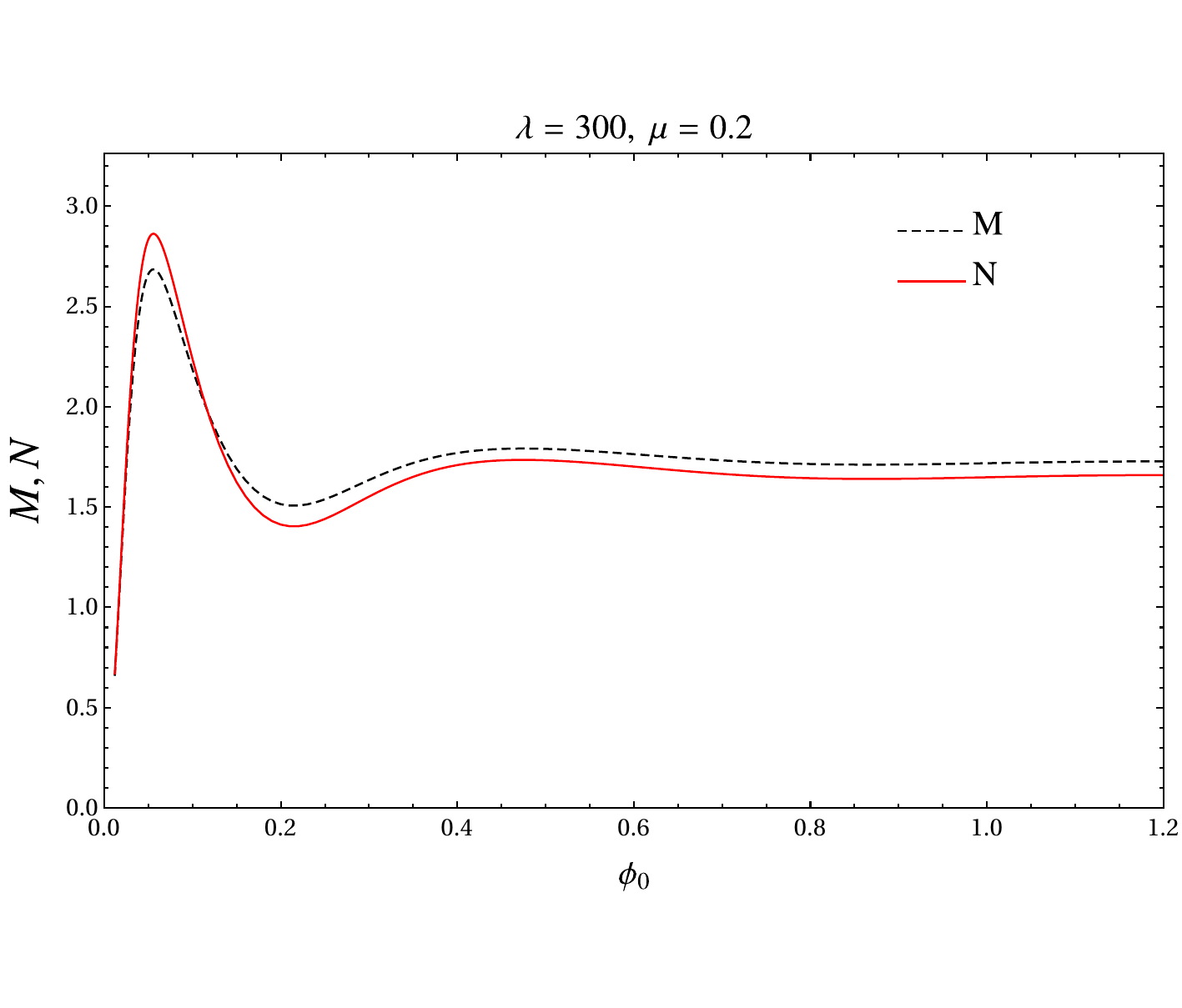}
}

\caption{The total mass $M$ and the number of particles $N$ for self-interacting E-FLS stars as a function of the  field at the origin $\phi_{0}$,  considering four different combinations of $\mu$ and $\lambda$. The regions where the number of particles $N$ is larger than the values of $M$ describe stable solutions.     
}
\label{mn_omega}
\end{figure*}

The relation between the ADM mass $M$ and the number of particles $N$ provides information about the  energetic stability of the self-interacting E-FLS stars. This information is given by the definition of binding energy~\cite{Friedberg1987a}:
\begin{align}
E_{b}=M-m\,N,
\end{align}
where $m$ is the mass of the particle that constitutes the scalar field. In this context,  when $M < m\,N$, the binding energy is negative, indicating that the system is gravitationally bound, since its total energy is lower than the sum of the individual particle rest masses, resulting in an energetically stable system. Otherwise, for $M > m\,N$, the binding energy is positive, meaning the configuration is gravitationally unbound and likely unstable against fission. In order to investigate the range of stable star solutions, we computed the number of particles for a family of self-interacting boson stars, considering four different combinations of the mass parameter $\mu$ and $\lambda$ solution using Eq.~\eqref{nnum}. In Fig.~\ref{mn_omega}, we present the number of particles and the ADM mass as functions of the complex scalar field at the origin  $\phi_{0}$ for different values of the mass parameter $\mu$ and self-interaction term $\lambda$. For $\lambda = 0$, the solutions with $\mu = 0$ exhibit a larger gap between the number of particles and the ADM mass, compared to those with $\mu = 0.2$ in the region $0 \leq \phi_{0} \lesssim 0.6$. When the self-interaction is introduced with $\lambda = 300$, the range of both the mass $M$ and the charge $N$ increases, altering their profiles relative to the case without self-interaction. In this case, the maximum values of $M$ and $N$ occur in the interval $0 \leq \phi_{0} \lesssim 0.2$. Furthermore, for $\mu = 0$ and $\lambda = 300$, the range of solutions with negative binding energy is notably larger, with $M < m\,N$ for $0 \leq \phi_{0} \lesssim 0.2$, reverting to negative binding energy again for $\phi_{0} \gtrsim 0.3$.



\begin{figure}
\includegraphics[scale=0.42]{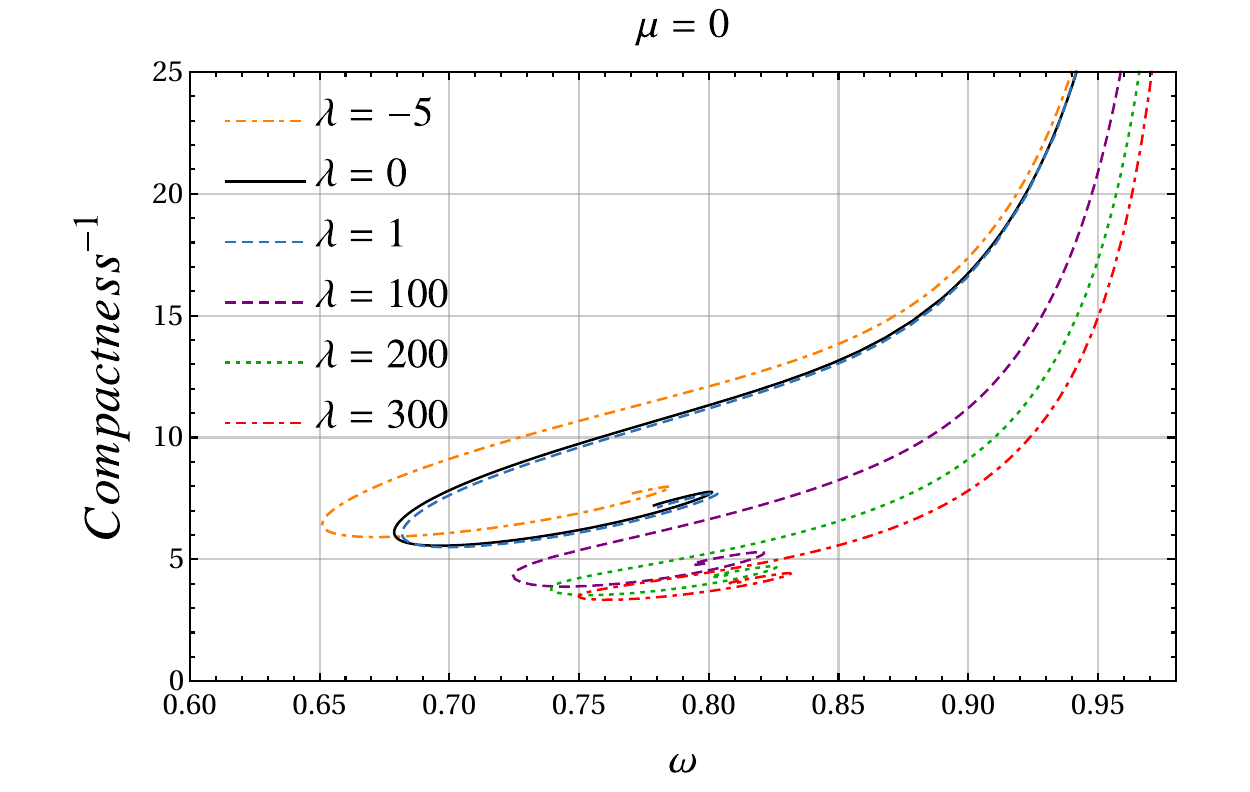}
\caption{The inverse compactness as a function of the oscillation frequency for distinct values of the self-interaction parameter $\lambda$ and $\mu=0$. As we increase the values of $\lambda$, we obtain a set of solutions that are more compact than the ones with smaller $\lambda$.}
\label{compac}
\end{figure}

We can also investigate the influence of the self-interaction parameter in the compactness of the E-FLS stars. To do so, we define the concept of inverse compactness, given by:
\begin{equation}
\text{Compactness}^{-1} \equiv \frac{R_\text{eff}}{2M_\text{eff}}, \label{compact}
\end{equation}
where $M_{\text{eff}}$ is the effective mass of the star that represents $99\%$ of its total ADM mass, i.e. $M_{\text{eff}}\equiv \mathcal{M}(R_{\text{eff}})$. The factor $1/2$ in Eq.~\eqref{compact} is suitable to compare the inverse compactness of BSs and black holes. Black holes have a well-defined radius, which corresponds to the radius of the event horizon with
$\text{Compactness}^{-1}\sim 1$, while less compact objects have larger inverse compactness. In Fig.~\ref{compac} we show the inverse compactness of the E-FLS stars, considering the mass parameter $\mu=0$ and varying the self-interaction term $\lambda$. In this figure, we notice that solutions with lower self-interaction parameter are less compact. For larger self-interaction strengths, the inverse compactness curve shifts downward, showing that E-FLS solutions become increasingly compact.

\section{Geodesic Analysis}
\label{sec:geo_ana}

To analyze the geodesic motion around BSs, we select a set of E-FLS solutions to explore numerically. These BS configurations are listed in Table \ref{efls_stars}, and the solutions EFLS1 to EFLS4 were selected due the fact that they correspond to the maximum mass star configuration for the given parameters $\mu$ and $\lambda$. These stars are not sufficiently compact to support the existence of light rings. To investigate the formation of light rings, we also explore the solutions EFLS5 and EFLS6, which are the most compact stars in Table~\ref{efls_stars}.


\begin{table}[htp]
\caption{The set of selected E-FLS star configurations and their properties.}
\centering
\label{efls_stars}
\begin{ruledtabular}
\begin{tabular}{c c c c c c c c}
Star & $\mu$ & $\lambda$ & $\omega$ & $M$ & $\phi_0$ & $r_{+}$ & $R_{\text{eff}}$ \\
\hline
EFLS1 & 0.2 & 0   & 0.8329 & 0.5700 & 0.24  &  - & 6.6796  \\
EFLS2 & 0.2 & 300 & 0.8733 & 2.6852 & 0.056 & - & 17.9915 \\
EFLS3 & 0   & 0   & 0.7139 & 0.5908 & 0.32  & - & 9.8835  \\
EFLS4 & 0   & 300 & 0.7688 & 2.8048 & 0.066 & - & 21.9458 \\
EFLS5 & 0   & 0   & 0.7819 & 0.3854 & 1.15  & 0.0840 & 5.6233  \\
EFLS6 & 0   & 300 & 0.8132 & 1.8377 & 1.15  & 0.0120 & 14.9198 \\
\end{tabular}
\end{ruledtabular}
\end{table}

\subsection{Timelike Geodesics in self-interacting E-FLS Stars}
\label{Subsec_timelike}

\begin{figure}[h!]
\includegraphics[scale=0.45]{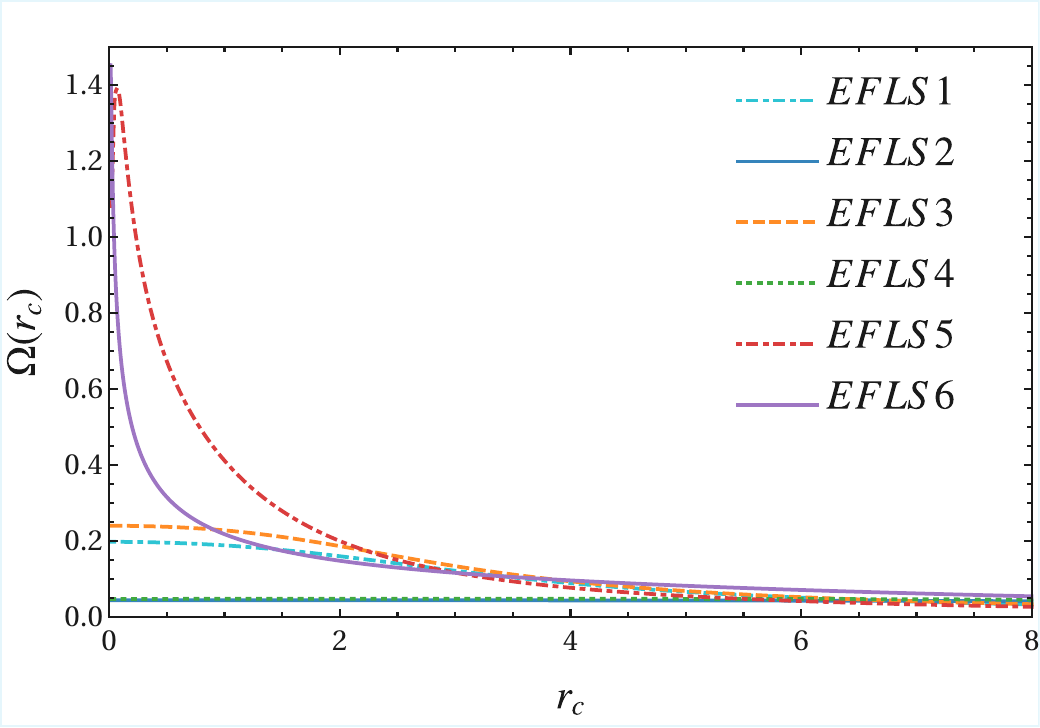}
\caption{Orbital frequency for timelike geodesics as a function of the radial coordinate $r$. We show six distinct configurations of massive E-FLS stars described in Table~\ref{efls_stars}. The more compact configurations are the ones with $\phi_{0}=1.15$, and they present highest orbital frequency.}
\label{Omega}
\end{figure}

The motion of massive particles around E-FLS stars is described by timelike geodesics. The corresponding Lagrangian is
\begin{align}
\label{lagran_timelike}2\mathcal{L}_t=g_{\mu\nu}\dot{x}^\mu\dot{x}^\nu=-e^\Gamma\,\dot{t}^2+e^\Lambda\,\dot{r}^2+r^2\dot{\varphi}^2=-1,
\end{align}
where we assume that the motion is restricted to the equatorial plane ($\theta=\pi/2$). The conserved quantities associated with time and angular symmetries are:
\begin{align}
\label{energy}&\bar{E}=e^\Gamma\,\dot{t}, \\
\label{momentum}&\bar{L}=r^2\dot{\varphi}.
\end{align}
The equations \eqref{energy} and \eqref{momentum} are referred to the energy $\bar{E}$ and angular momentum $\bar{L}$ of the particle. Substituting Eqs.~\eqref{energy} and \eqref{momentum} in Eq.~\eqref{lagran_timelike}, we obtain a radial equation:
\begin{align}
\label{rad_eq_timelike}e^{\left( \Gamma+\Lambda\right)}\,\dot{r}^2=\bar{E}^2-e^\Gamma\left(1+\frac{\bar{L}^2}{r^2}\right).
\end{align}
The radius where the particles follow a circular orbit is defined as $r_c$. At $r_c$, the circular orbits satisfy $\dot{r} = 0$ and $\ddot{r} = 0$, yielding:
\begin{align}
\label{E-L-timelike}&\bar{E}_c^2=\left. \frac{2\,e^\Gamma}{2-r\,\Gamma'}\right|_{r=r_c}, \quad \bar{L}_c^2=\left. \frac{r^3\,\Gamma'}{2-r\,\Gamma'} \right|_{r=r_c}.
\end{align}
The orbital frequency quantifies the rate at which a particle completes its revolution around the star as measured by a distant observer. It is defined as the time derivative of the azimuthal coordinate
\begin{equation}
\Omega = \frac{d\phi}{dt}.
\end{equation}
Evaluating the orbital frequency for massive particles provides insight into the dynamics of the accretion disk associated with the E-FLS star configurations, such as its radial extent. The orbital frequency at the critical radius $r_c$ is given by:
\begin{align}
\label{Omega-timelike}\Omega(r_c)=\left. \left[\frac{e^\Gamma\,\Gamma'}{2\,r}\right]^\frac{1}{2} \right|_{r=r_c}.
\end{align}
Figure~\ref{Omega} illustrates $\Omega(r_c)$ for the configurations in Table~\ref{efls_stars}. For large $r$, the behavior remains similar for all solutions. However, near the center, EFLS5 and EFLS6 exhibit higher orbital frequencies. In the case of EFLS6, the contribution from the positive self-interaction term leads to the appearance of a maximum closer to the origin compared to the EFLS5 solution and it is also possible to notice that massive particles in EFLS6 exhibit higher orbital frequencies on unstable orbits. Comparing the EFLS5 and EFLS6 solutions, it is possible to see that the orbital frequency, as we approach the origin, is larger for the self-interacting case. In both configurations, massive particles can circularly orbit very close to the origin.    

\subsection{Null geodesics in self-interacting E-FLS Stars}
\label{Null_geo_sec}

Null geodesics describe the paths of light in a curved spacetime and are crucial for correctly interpret the images of compact objects~\cite{Universal_Int, Li2025, Pringle1981}. The Lagrangian governing null geodesics in E-FLS stars is:
\begin{align}
\label{Lagran}
2\,\mathcal{L}_{ph} = g_{\mu\nu}\dot{x}^\mu\dot{x}^\nu = -e^\Gamma \dot{t}^2 + e^\Lambda \dot{r}^2 + r^2\dot{\varphi}^2 = 0.
\end{align}
As in the timelike geodesics case, we set $\theta = \pi/2$ without loss of generality. The null geodesics also admits two conserved quantities
\begin{align}
\label{energy_photons_eq}E=e^\Gamma\,\dot{t},\\
\label{angular_photons_eq}L=r^2\dot{\phi},
\end{align}
related to the energy and angular momentum of photons, respectively. Substituting Eqs.~\eqref{energy_photons_eq} and \eqref{angular_photons_eq} into Eq.~\eqref{Lagran}, we obtain the radial equation of motion for photons
\begin{align}
\label{rad_eq}
\dot{r}^2 + V(r, E, L) = 0,
\end{align}
where the term $V(r, E, L)$ is the effective potential governing photon motion, given by
\begin{align}
\label{ef_pot}
V(r, L, E) = e^{-\Lambda} \left(\frac{L^2}{r^2} - \frac{E^2}{e^{\Gamma}}\right).
\end{align}
The circular photon orbits, aka light-rings, occur at the radii $r_{c}$. At this radius coordinate, the effective potential behaves as:
\begin{align}
\label{LR1}
V(r_{c}, L, E) = 0, \quad V'(r_{c}, L, E) = 0.
\end{align}
Since $V(r, L, E)$ depends on the constants of motion, it is convenient to define a novel effective potential independent of $E$ and $L$, given by: 
\begin{align}
\label{H_pot}
\mathcal{H}(r) \equiv \frac{e^{\frac{\Gamma}{2}}}{r}. 
\end{align}
Given the definition in Eq. \eqref{H_pot}, we can now express the potential $V(r, L, E)$ as:
\begin{align}
V(r, L, E) = \frac{L^2}{e^{\Gamma+\Lambda}} \left(\mathcal{H}(r) + \frac{1}{b} \right) \left(\mathcal{H}(r) - \frac{1}{b} \right),
\label{newve}
\end{align}
which reveal the conditions for the existence of circular photon orbits. In order to the equation \eqref{newve} satisfy the conditions in the Eq. \eqref{LR1}, the function $\mathcal{H}(r)$ must obey:
\begin{align}
\label{LR_Conds}
\mathcal{H}(r_{c}) = \frac{1}{b}, \quad \mathcal{H}'(r_{c}) = 0.
\end{align}

\begin{figure}
\includegraphics[scale=0.3]{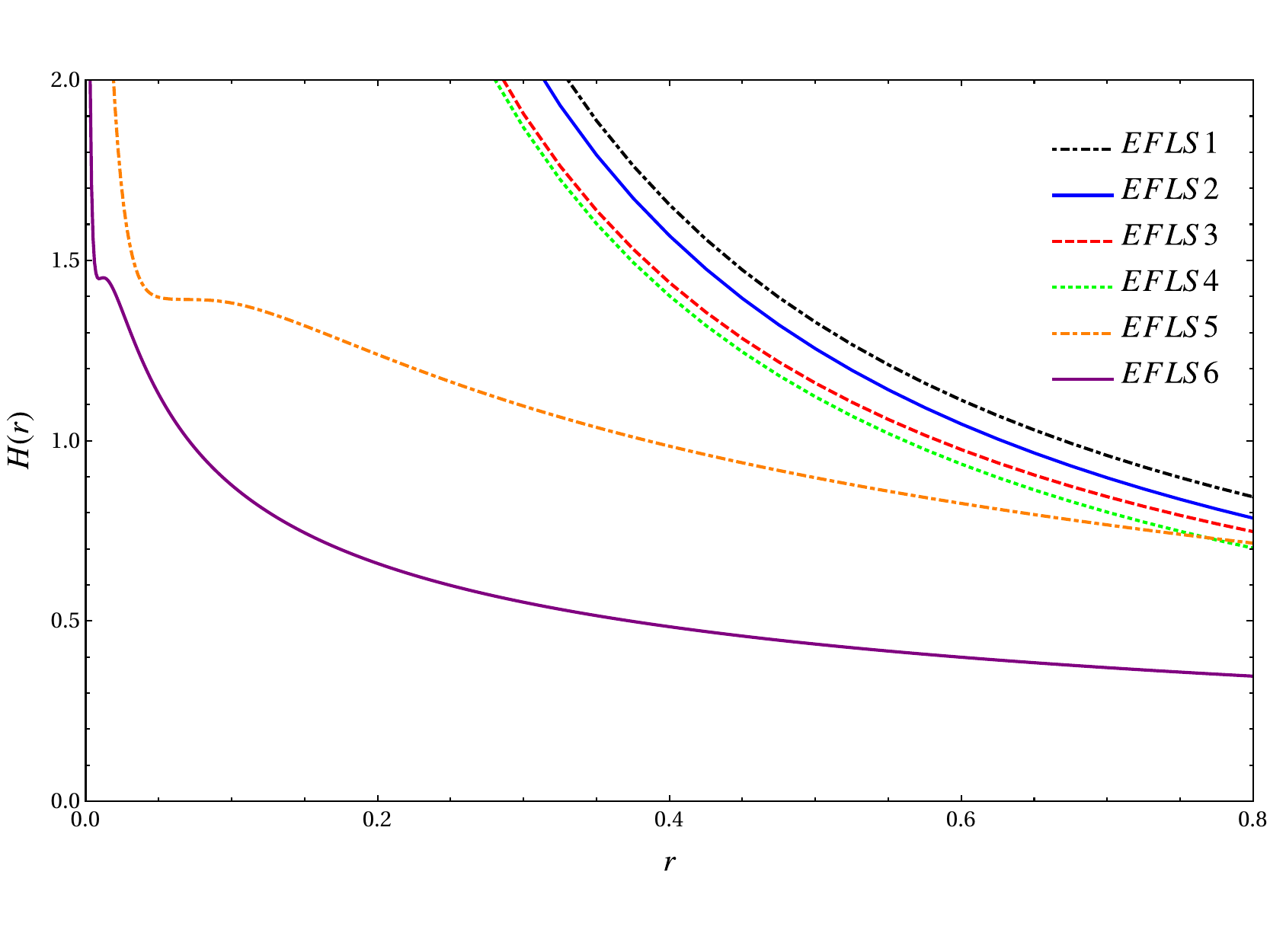}
\caption{The effective potential $\mathcal{H}(r)$ for null geodesics in self-interacting E-FLS star spacetimes, as a function of the radial coordinate, considering the different configurations presented in Table~\ref{efls_stars}.}
\label{h_pot}
\end{figure}

Figure~\ref{h_pot} shows $\mathcal{H}(r)$ for different configurations in Table~\ref{efls_stars}. Although all the solutions in the range EFLS1 to EFLS4 are the ones with maximum-mass for the chosen parameters, they do not admit circular photon orbits: the null effective potential $\mathcal{H}(r)$ shows no stationary points for these cases. This result is consistent with our selection of the maximum-mass branch for each pair $(\mu,\lambda)$. In the non-self-interacting limit, maximal-mass configurations are typically not compact enough to support light rings~\cite{Cunha_Lensing_Dynamics}. On the other hand, EFLS5 and EFLS6 are compact enough for $\mathcal{H}(r)$ to develop extrema, and consequently admit light rings. It is well known that, for horizonless compact solutions, the light rings appear in pairs. The inner one ($r_{-}$) is a stable light ring and the outer one ($r_{+}$) is an unstable light ring~\cite{Cunha2017, Cunha2020}. Comparing EFLS5 with EFLS6, the light rings radius $r_{\pm}$ are larger for EFLS5. Increasing the self-interaction parameter $\lambda$ enhances compactness of the stars, which in turn shifts diminishes the light rings radii. Hence $r_{\pm}$ decreases as we increase $\lambda$.

\begin{figure*}
  \centering
  \subfloat[EFLS1]{\includegraphics[scale=0.25]{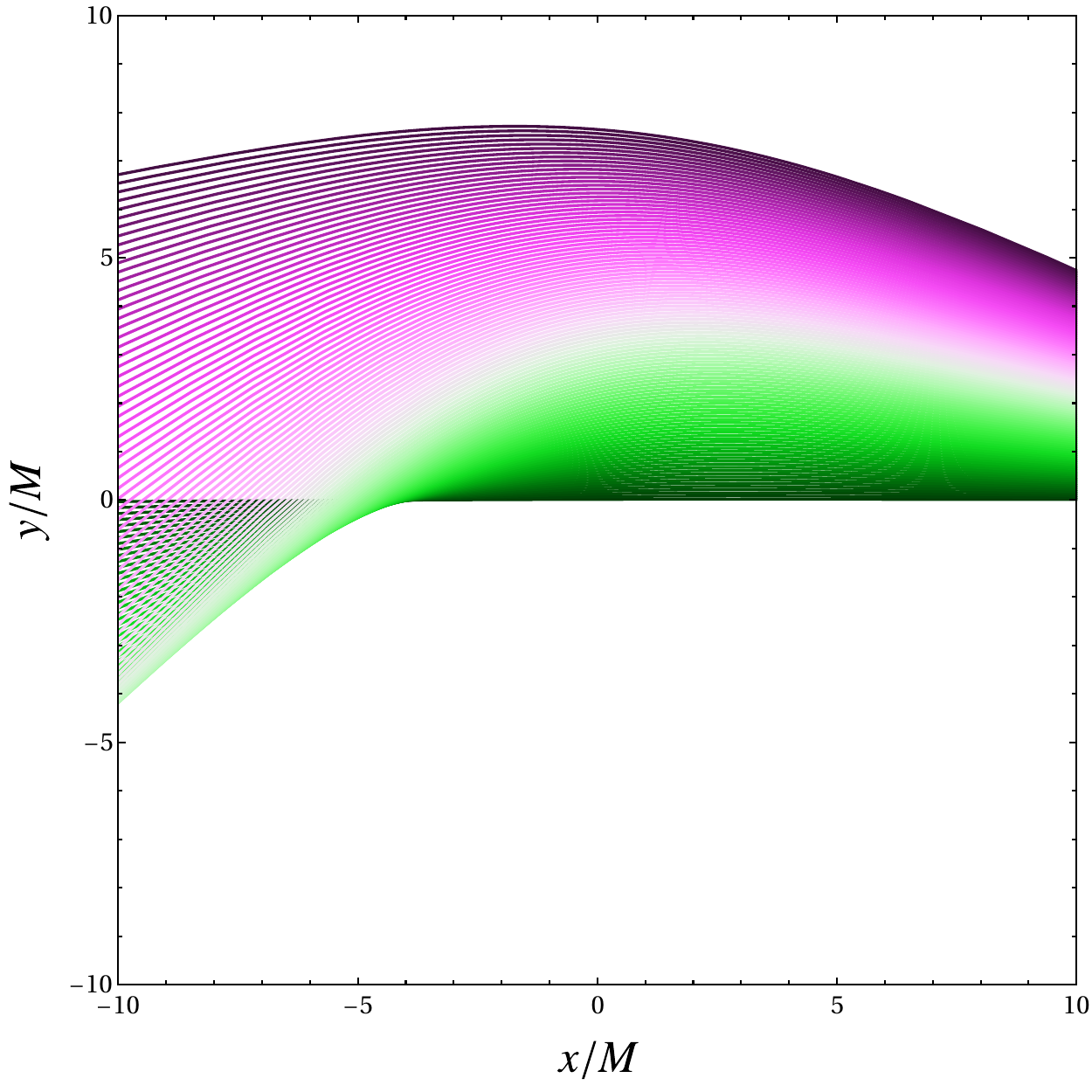}\label{ageo}}
  \subfloat[EFLS2]{\includegraphics[scale=0.25]{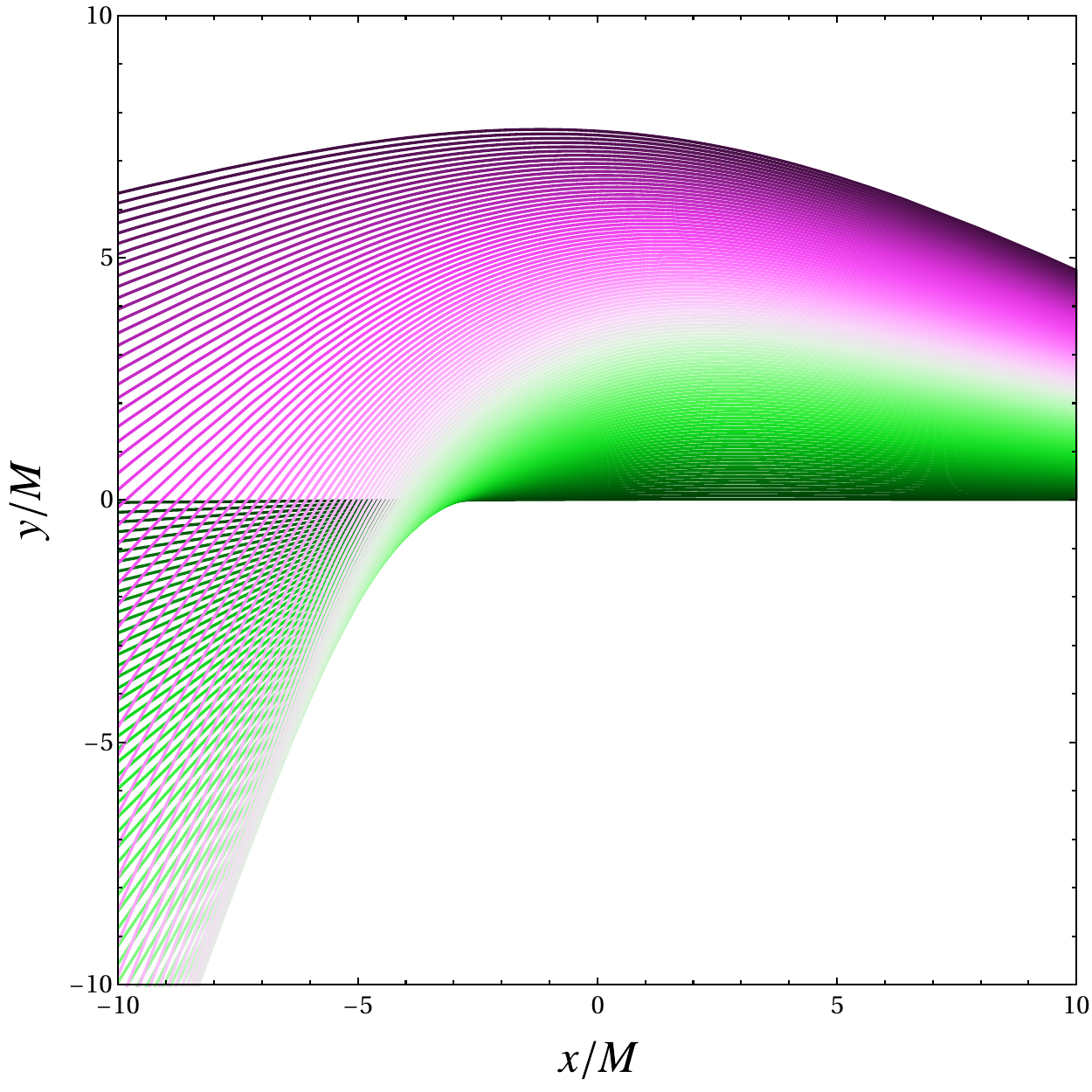}\label{bgeo}}
  \subfloat[EFLS3]{\includegraphics[scale=0.25]{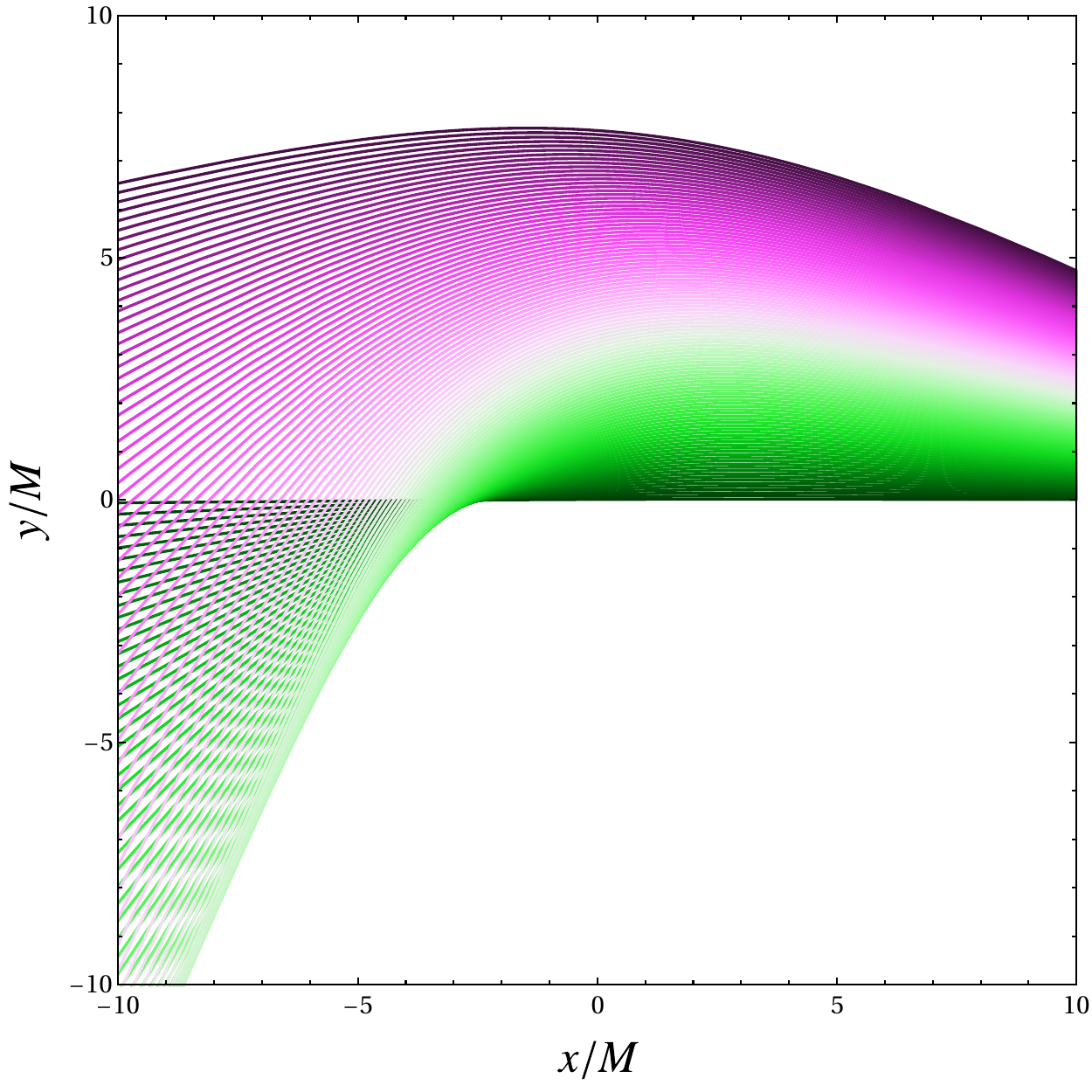}\label{cgeo}}
    \\
  \subfloat[EFLS4]{\includegraphics[scale=0.25]{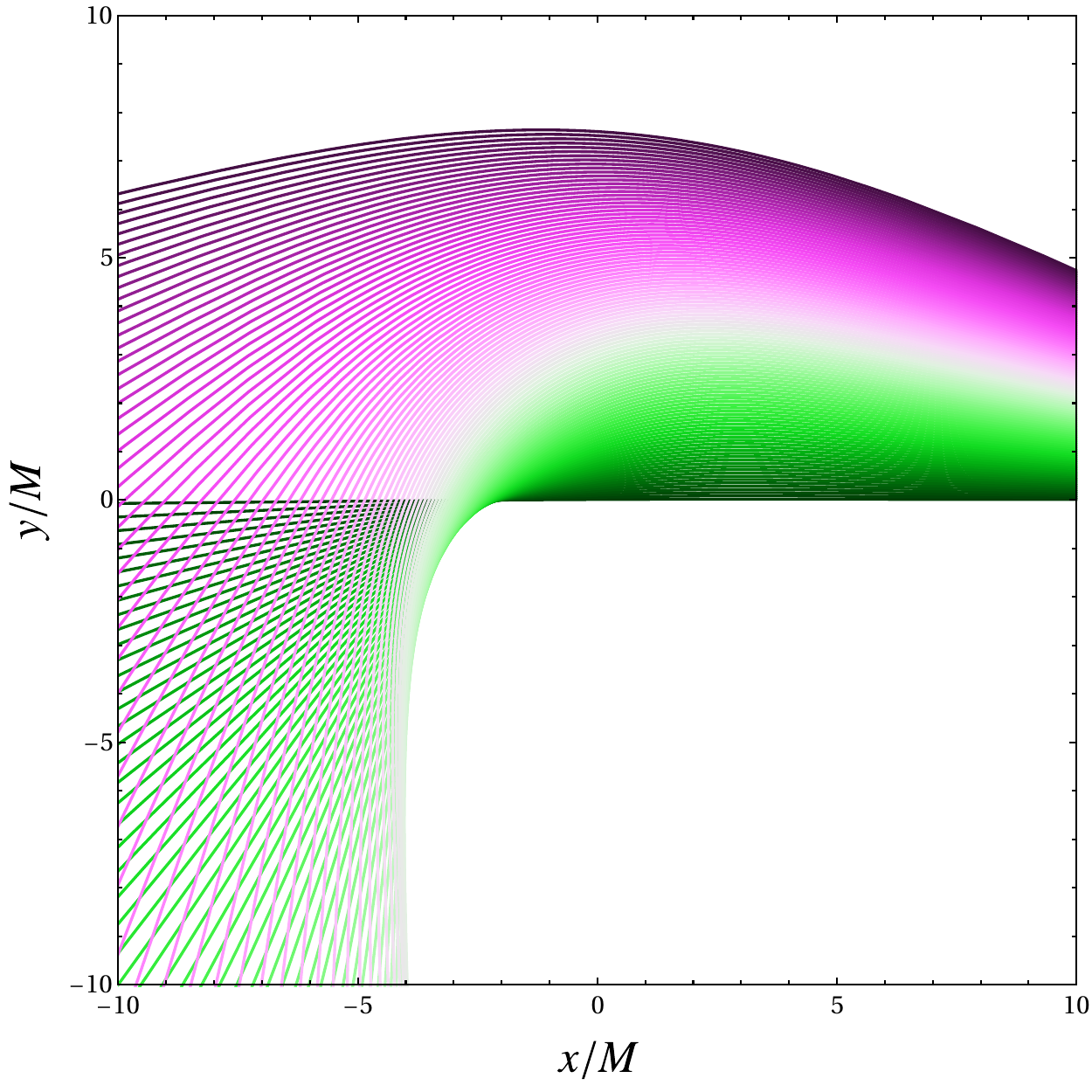}\label{dgeo}} 
  \subfloat[EFLS5]{\includegraphics[scale=0.25]{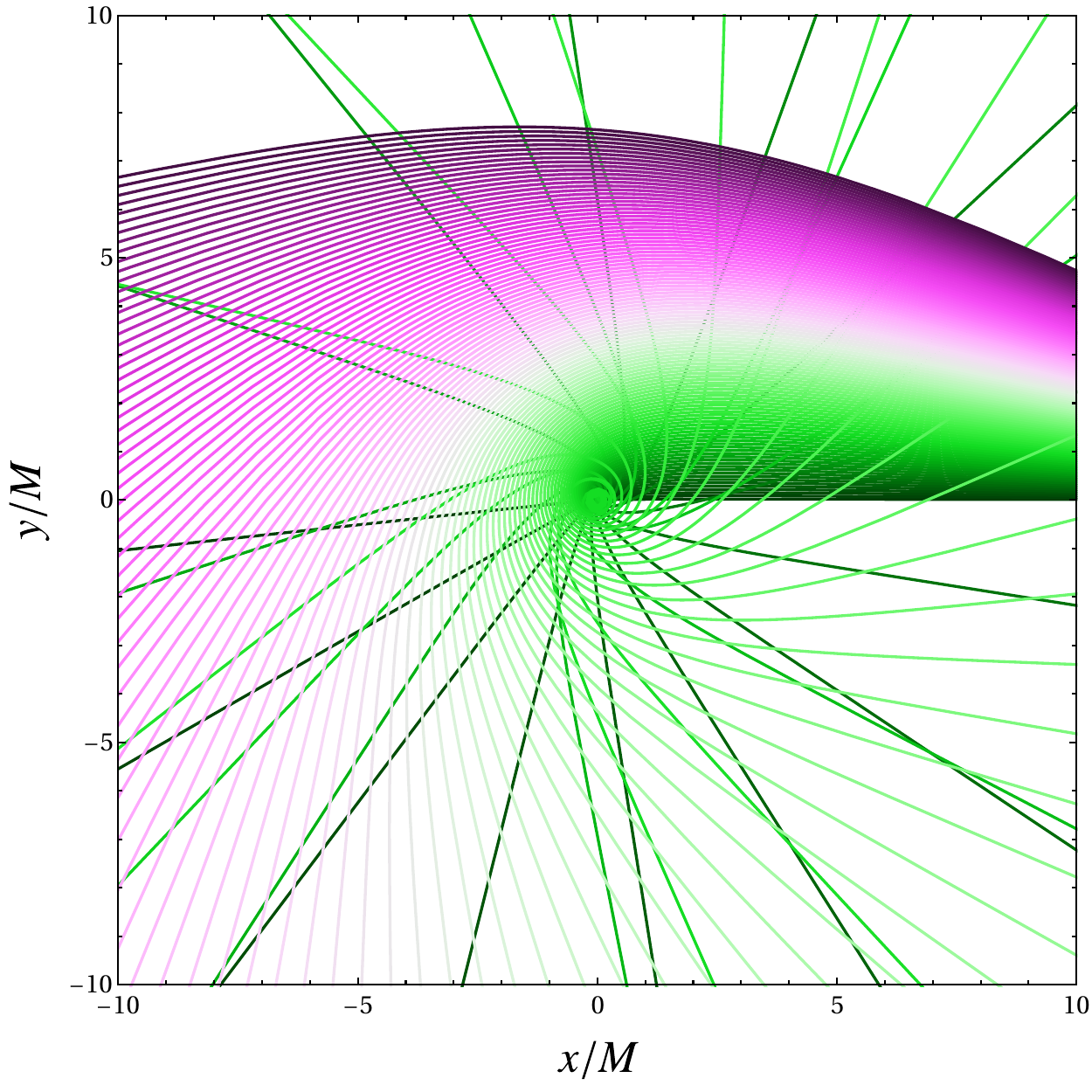}\label{egeo}}
  \subfloat[EFLS6]{\includegraphics[scale=0.25]{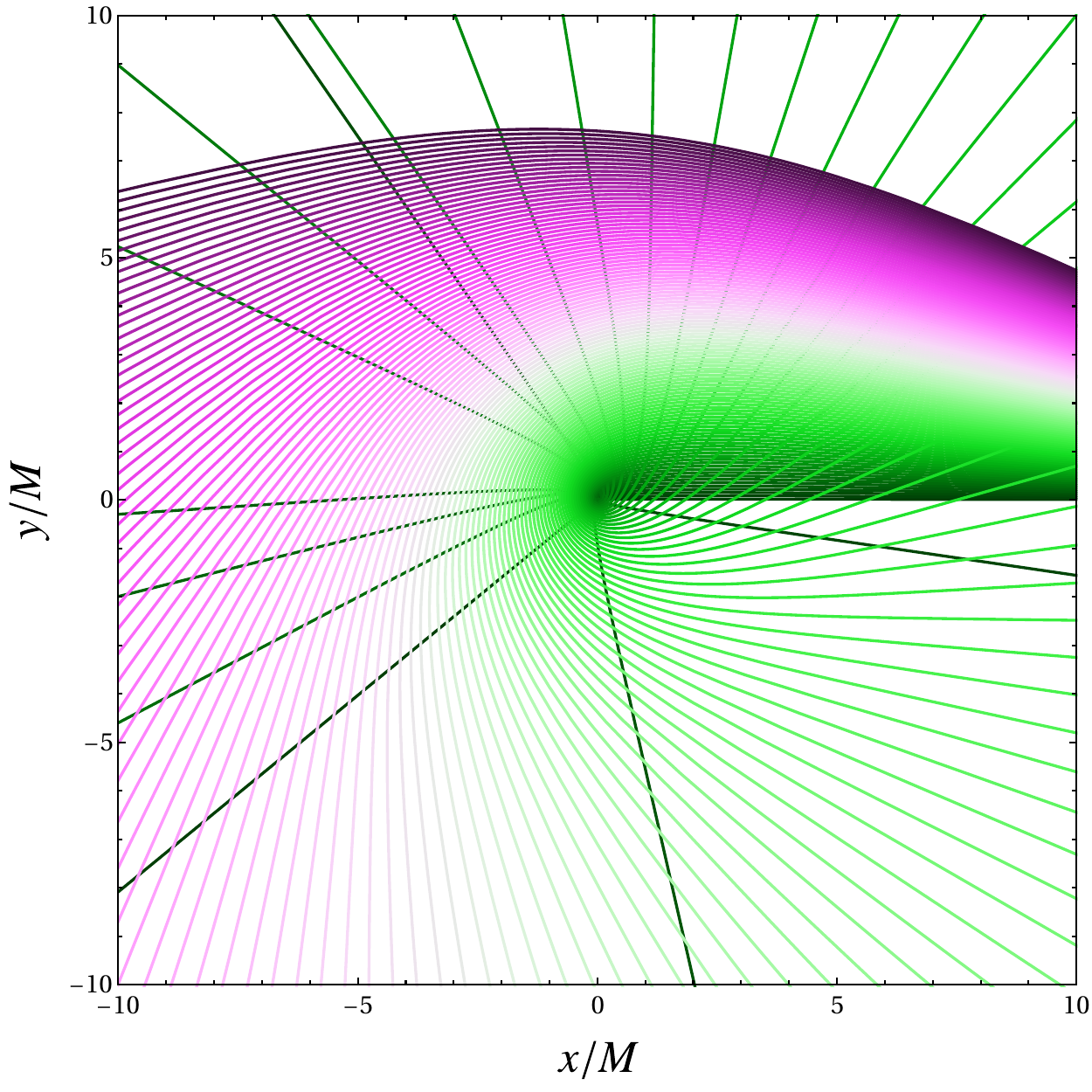}\label{fgeo}}
\caption{The trajectories described by null geodesics with different impact parameters along the equatorial plane of the self-interacting E-FLS stars. The color scheme represents the different impact parameters for null geodesics. }
\label{geo_path}
\end{figure*}

We show the null geodesic paths around the E-FLS stars given in Table ~\ref{efls_stars}. The results considering a range of distinct impact parameters is shown in Fig.~\ref{geo_path}. Although configurations EFLS1 to EFLS4 correspond to solutions with maximum mass values, they do not present closed null geodesic orbits.
As shown in subfigures~\ref{ageo}–\ref{dgeo}, the trajectories of null geodesics are modified by the presence of the star, located at the center of the plot. However, the resulting deflection is not sufficient to produce circular orbits. This behavior changes for the solutions displayed in subfigures~\ref{egeo} and~\ref{fgeo}, which are compact enough to bend the null geodesics and give rise to closed circular orbits.

\section{Astrophysical Images of self-interacting E-FLS Stars}
\label{sec:astro_images}

We consider four different configurations of E-FLS stars, two of which do not exhibit circular photon orbits, while the other two allow for circular photon orbits. The existence of circular photon orbits influences the bending of light and thus affects the appearance of the E-FLS star's shadow as seen by a distant observer. The procedure to obtain the shadow images is the same as used in Ref.~\cite{Harolo2021}. We use a backward ray-tracing method, propagating light rays from the observer until they either intersect the accretion disk or escape to infinity. The path described by light is governed by the first-order null geodesic equations \cite{Cunha2018}:
\begin{align}
\label{energy_1} \dot{t}=\frac{E}{e^\Gamma},  \\
\label{momentum_phi} \dot{\phi}=\frac{L}{r^2},
\end{align}
and by the second-order geodesic equations for the radial and polar coordinates:
\begin{align}
\label{dotr}\ddot{r}+\Gamma^r_{\ \mu\nu}\dot{x}^\mu\dot{x}^\nu=0,\\
\label{dottheta}\ddot{\theta}+\Gamma^{\theta}_{\ \mu\nu}\dot{x}^\mu\dot{x}^\nu=0,
\end{align}
where $\Gamma^\alpha_{\ \mu\nu}$ denotes the Christoffel symbols of the E-FLS star geometry. In order to simulate the E-FLS stars' images surrounded by an emitting accretion disk, the geodesic equations must be solved coupled with the radiative transfer equation. We assume that the accretion disk emits unpolarized radiation. Under this assumption, the radiative transfer equation takes the form given in \cite{Lindquist}.
\begin{align}
\label{transfer_eq}\frac{d}{d\lambda}\left(\frac{I_\nu}{\nu^3} \right)=\frac{j_\nu}{\nu^2} - \nu\,\alpha_\nu\,\left(\frac{I_\nu}{\nu^3} \right)
\end{align}
In equation~\eqref{transfer_eq}, $I_{\nu}$, $j_{\nu}$ and $\alpha_{\nu}$ are, respectively, the specific intensity, the emission coefficient and the absorption coefficient:
\begin{align}
\label{invar_quat}\mathcal{I}=\frac{I_\nu}{\nu^3}, \quad \eta=\frac{j_\nu}{\nu^2}, \quad \chi=\nu\,\alpha_\nu.
\end{align}

Since the light rays are evolved backward in time from the observer towards the star, we impose as part of the initial conditions: 
\begin{equation}
t=0,\quad r=r_{obs},\quad \theta=\theta_{obs},\quad \varphi=0.
\end{equation}
We must determine the initial light ray's direction as measured by a static local observer. To do this, we introduce a local orthonormal tetrad for a static local observer:
\begin{eqnarray}
&&\hat{\lambda}^{\hat{0}}_{\ \mu}=\left(e^{\frac{\Gamma}{2}} ,\ 0,\ 0,\ 0 \right), \label{t1} \\
&&\hat{\lambda}^{\hat{1}}_{\ \mu}=\left(0, \ e^{\frac{\Lambda}{2}} ,\ 0,\ 0 \right), \label{t2}\\
&&\hat{\lambda}^{\hat{2}}_{\ \mu}=\left(0 ,\ 0,\ r,\ 0 \right), \label{t3}\\
&&\hat{\lambda}^{\hat{3}}_{\ \mu}=\left(0 ,\ 0,\ 0,r\sin\theta \label{t4} \right),
\end{eqnarray}
where the hats specify the tetrad indices. We can project the four-momentum into the tetrad basis~\eqref{t1}-\eqref{t4}, using
\begin{equation}
p_{\mu} = p_{\hat{a}}\hat{\lambda}^{\hat{a}}_{\ \mu}, \label{4m}
\end{equation}
where $p_{\hat{a}}$ is the four-momentum of the photon as measured by the local static observer. We can parametrize $p_{\hat{a}}$ using a pair of angles $(\alpha, \beta)$ defined in the local observer frame as
\begin{align}
\label{pt_frame}&p_{\hat{r}}=|p|\cos\alpha\cos\beta,\\
&p_{\hat{\phi}}=|p|\cos\alpha\sin\beta,\\
\label{ptheta_frame}&p_{\hat{\theta}}=|p|\sin\alpha.
\end{align}
Given a pair of angles $(\alpha, \beta)$, Eqs.~\eqref{pt_frame}-\eqref{ptheta_frame} define the remaining initial conditions for the backwards ray-tracing. Solving the geodesic and radiative transfer equations for a grid of initial directions yields a map $(\alpha, \beta) \mapsto I^{obs}(x,y)$, where $(x,y)$ are image plane coordinates and $I^{obs}(x,y)$ the corresponding observed intensity.

We have selected four configurations of Table~\ref{efls_stars} to perform the backward ray-tracing simulations. In the following subsections, we present the astrophysical images for these configurations. To investigate the contribution of the accretion disk to the shadow image, we consider two different accretion disk models surrounding these stars. Given that all the star solutions are numerical, the results were obtained by implementing the cubic spline interpolation of such solutions in the ray-tracing computational routine.


\begin{figure*}
\subfloat[EFLS3]{
\includegraphics[scale=0.45]{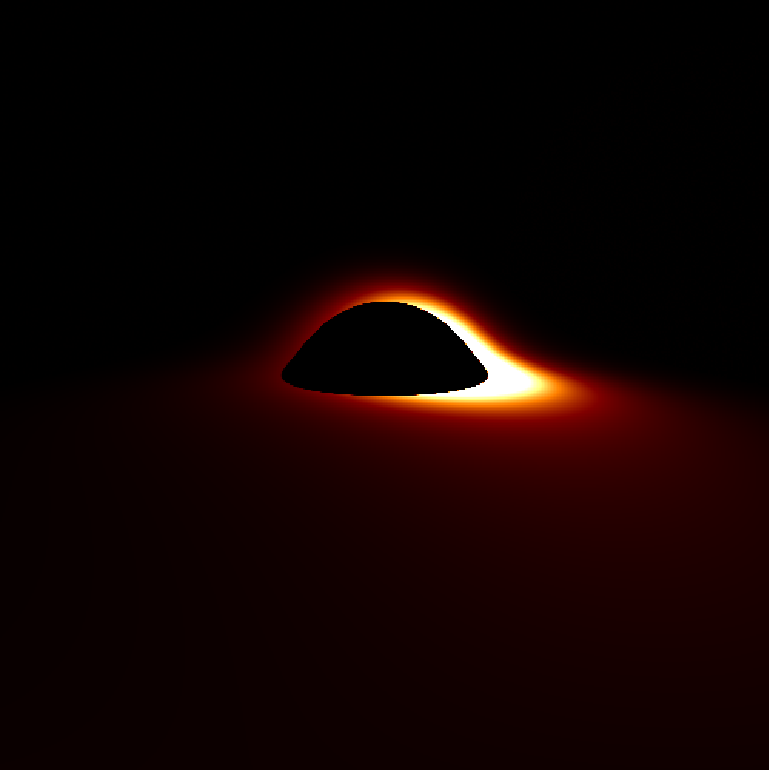}}
\subfloat[EFLS4]{
\includegraphics[scale=0.45]{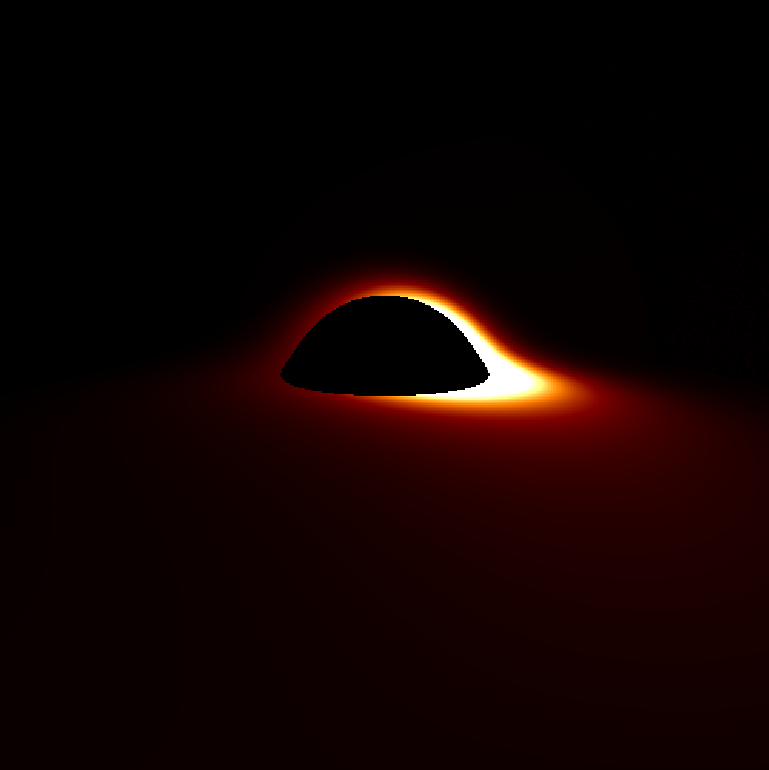}}
\\
\subfloat[EFLS5]{
\includegraphics[scale=0.45]{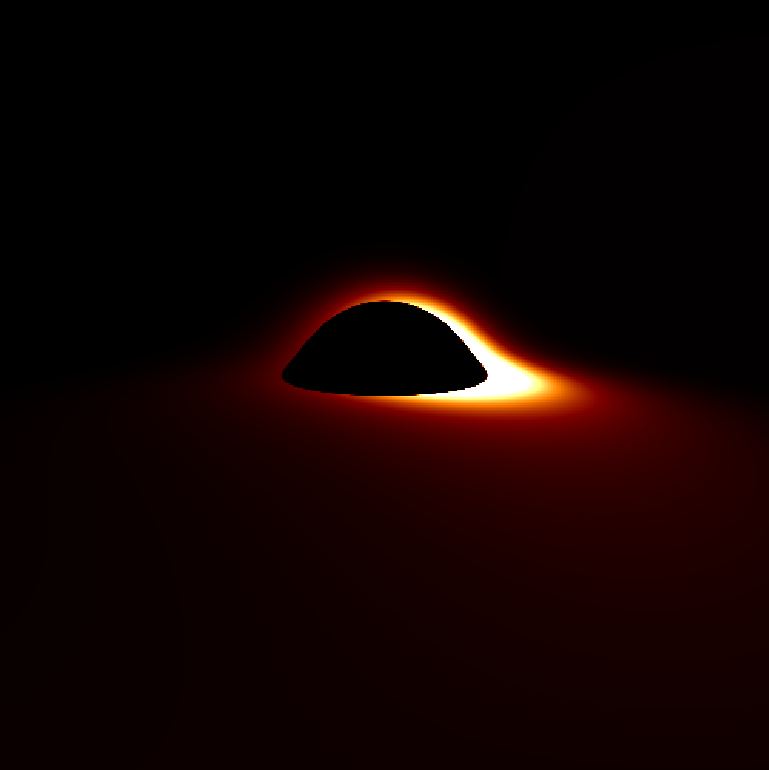}}
\subfloat[EFLS6]{
\includegraphics[scale=0.45]{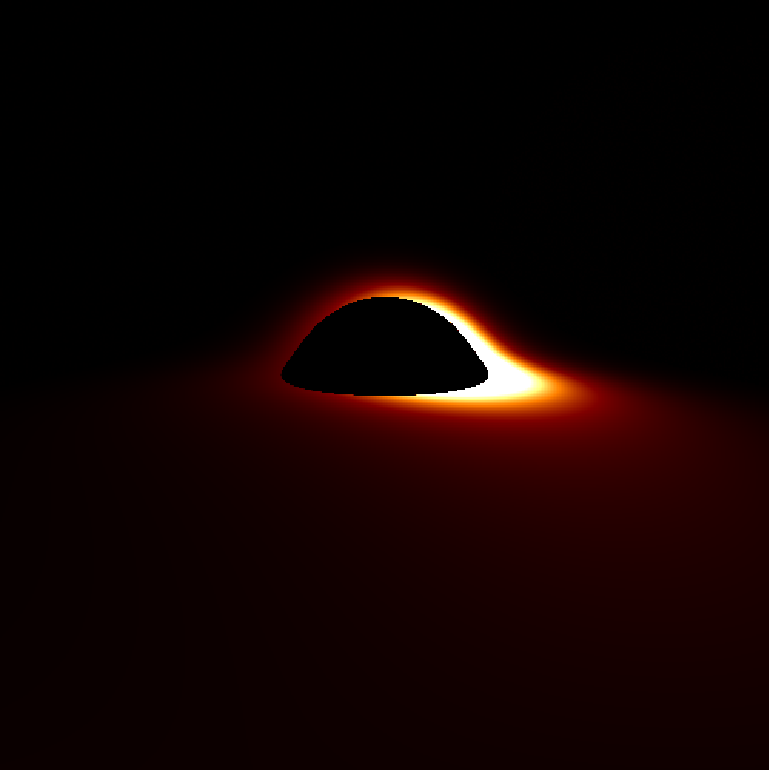}}

\caption{Intensity map of some E-FLS stars surrounded by an optically thick accretion disk. The observer location is $r_{obs} = 20 M $ and $\theta_{\textrm{obs}}=80^\circ$.}
    \label{Lensing_EFLS}
\end{figure*}

\begin{figure*}
\subfloat[EFLS3]{
\includegraphics[scale=0.45]{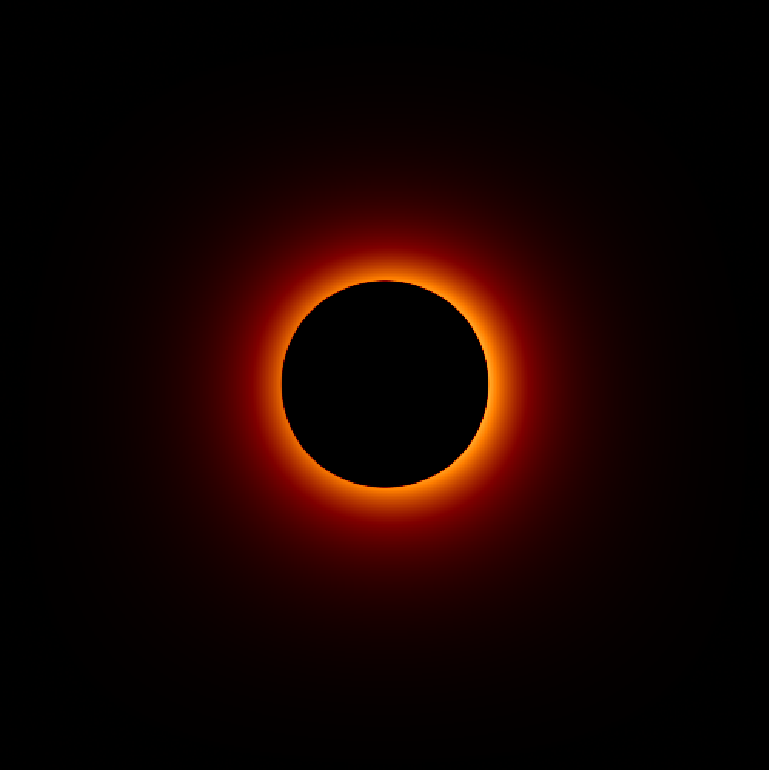}}
\subfloat[EFLS4]{
\includegraphics[scale=0.45]{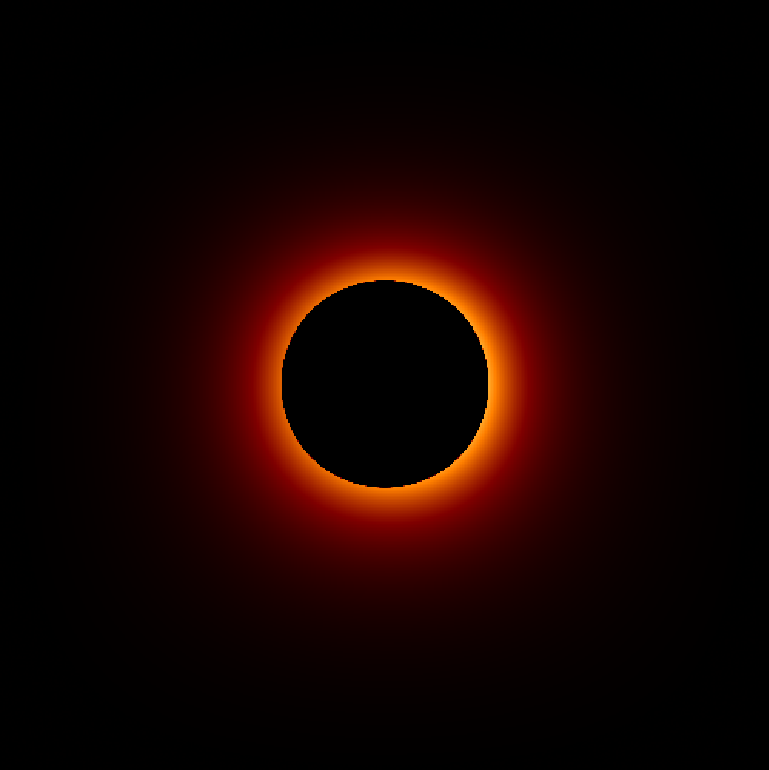}}
\\
\subfloat[EFLS5]{
\includegraphics[scale=0.45]{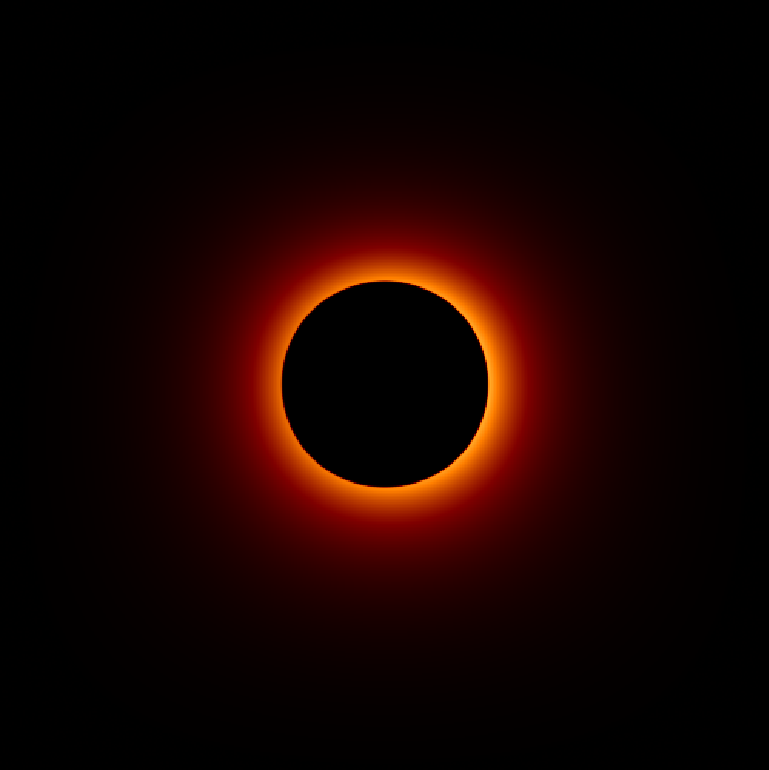}}
\subfloat[EFLS6]{
\includegraphics[scale=0.45]{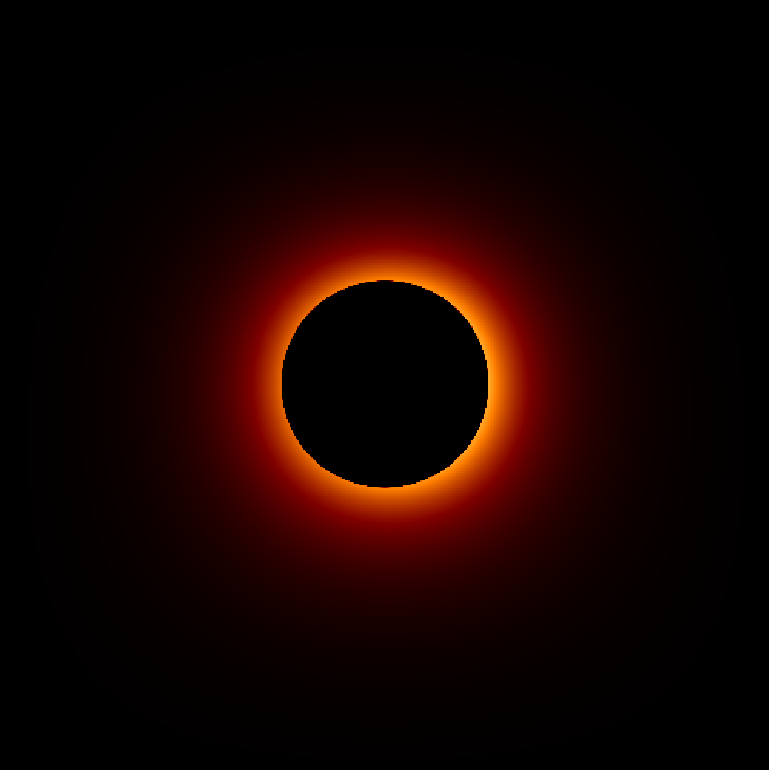}}

\caption{Intensity map of some E-FLS stars surrounded by an optically thick accretion disk. The observer location in this simulation is $r_{obs} = 20 M $ and $\theta_{\textrm{obs}}=5^\circ$.}
    \label{Lensing_EFLS2}
\end{figure*}

\begin{figure*}
\subfloat[EFLS3]{
\includegraphics[scale=0.45]{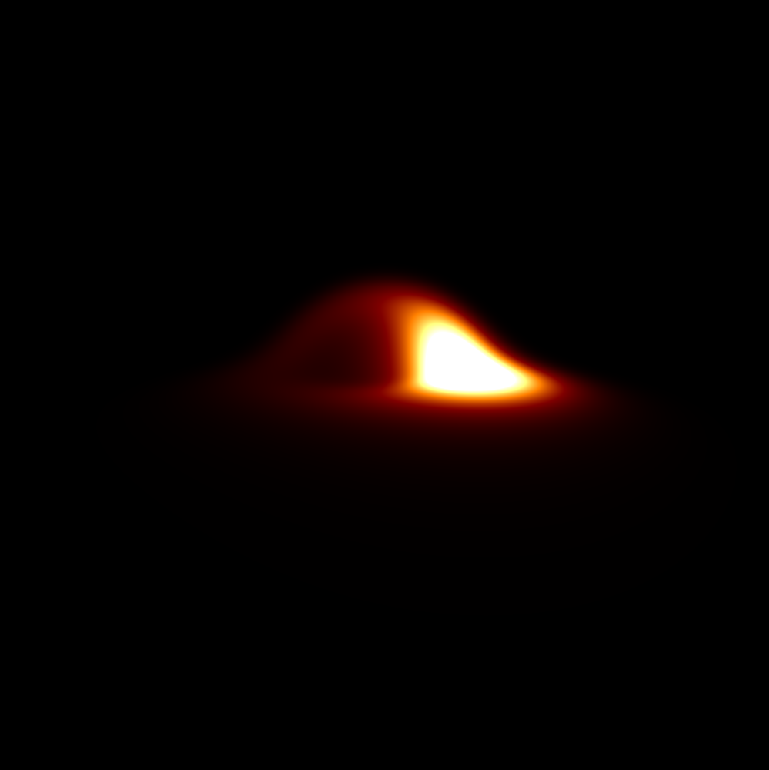}}
\subfloat[EFLS4]{
\includegraphics[scale=0.45]{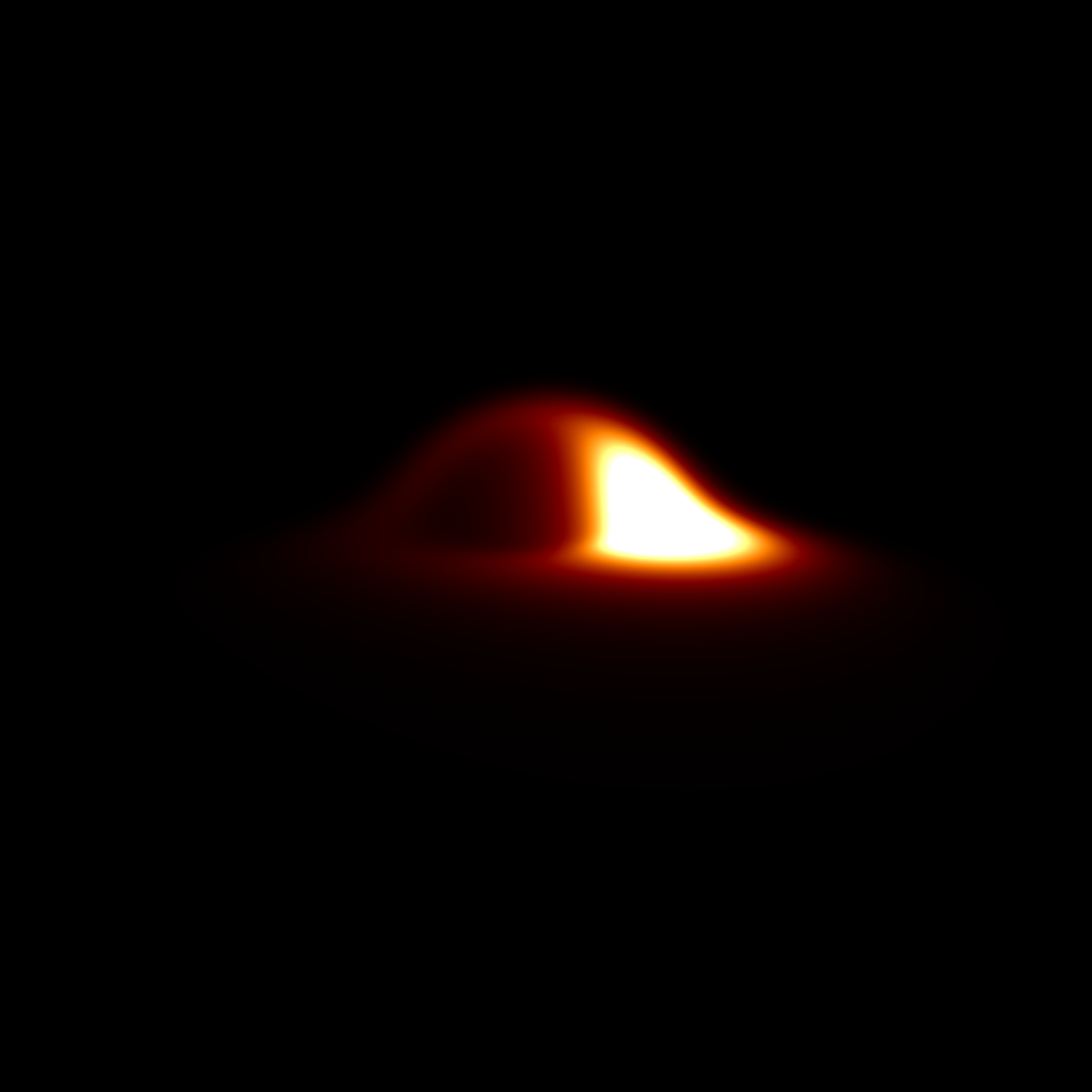}}
\\
\subfloat[EFLS5]{
\includegraphics[scale=0.45]{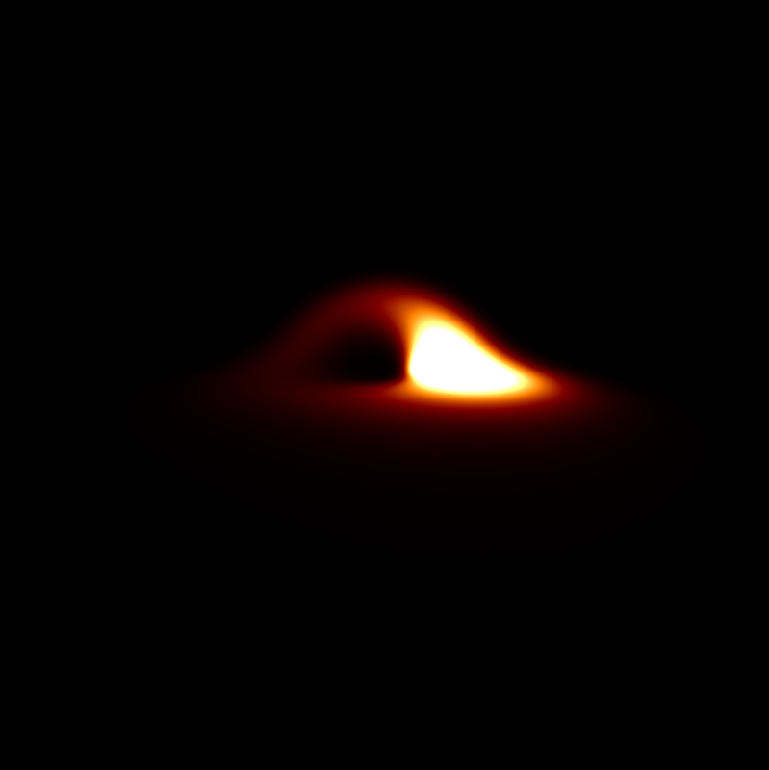}}
\subfloat[EFLS6]{
\includegraphics[scale=0.45]{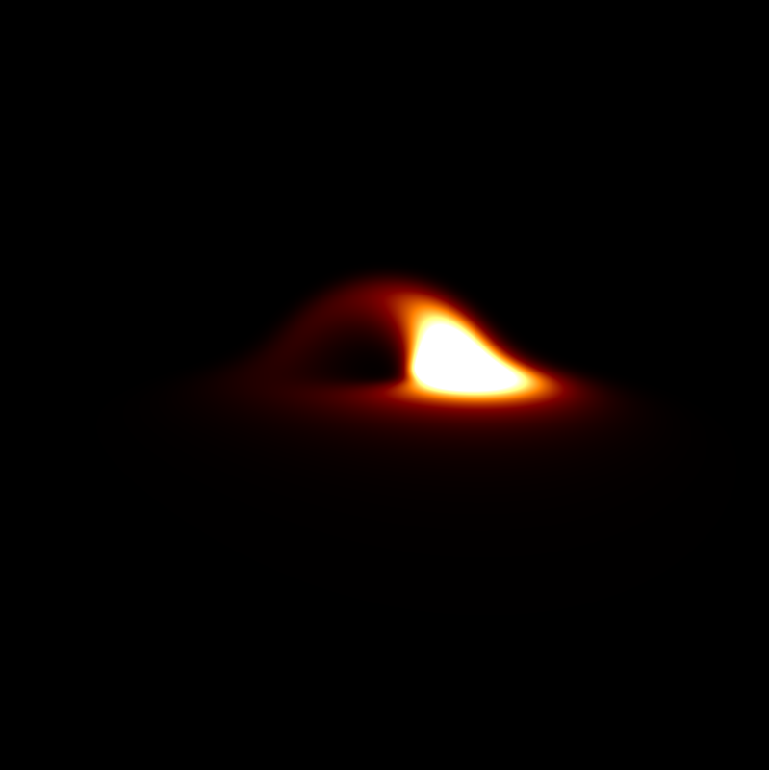}}

\caption{Intensity map of some E-FLS stars surrounded by an optically thick accretion disk, with no inner edge, extending until the origin. The observer location is $r_{obs} = 20 M $ and $\theta_{\textrm{obs}}=80^\circ$.}
    \label{newdisk_Lensing_EFLS}
\end{figure*}

\begin{figure*}
\subfloat[EFLS3]{
\includegraphics[scale=0.94]{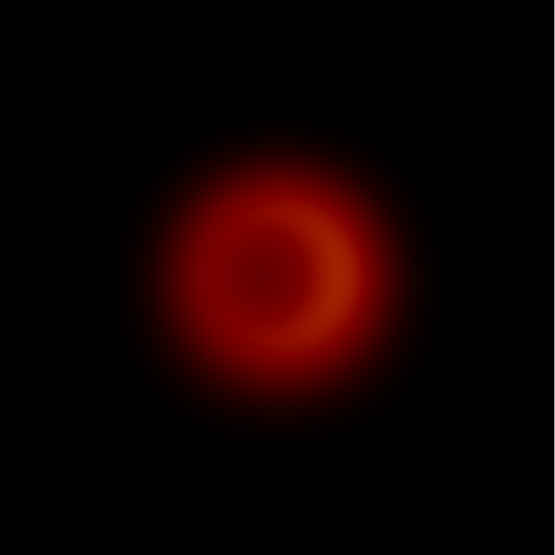}}
\subfloat[EFLS4]{
\includegraphics[scale=0.45]{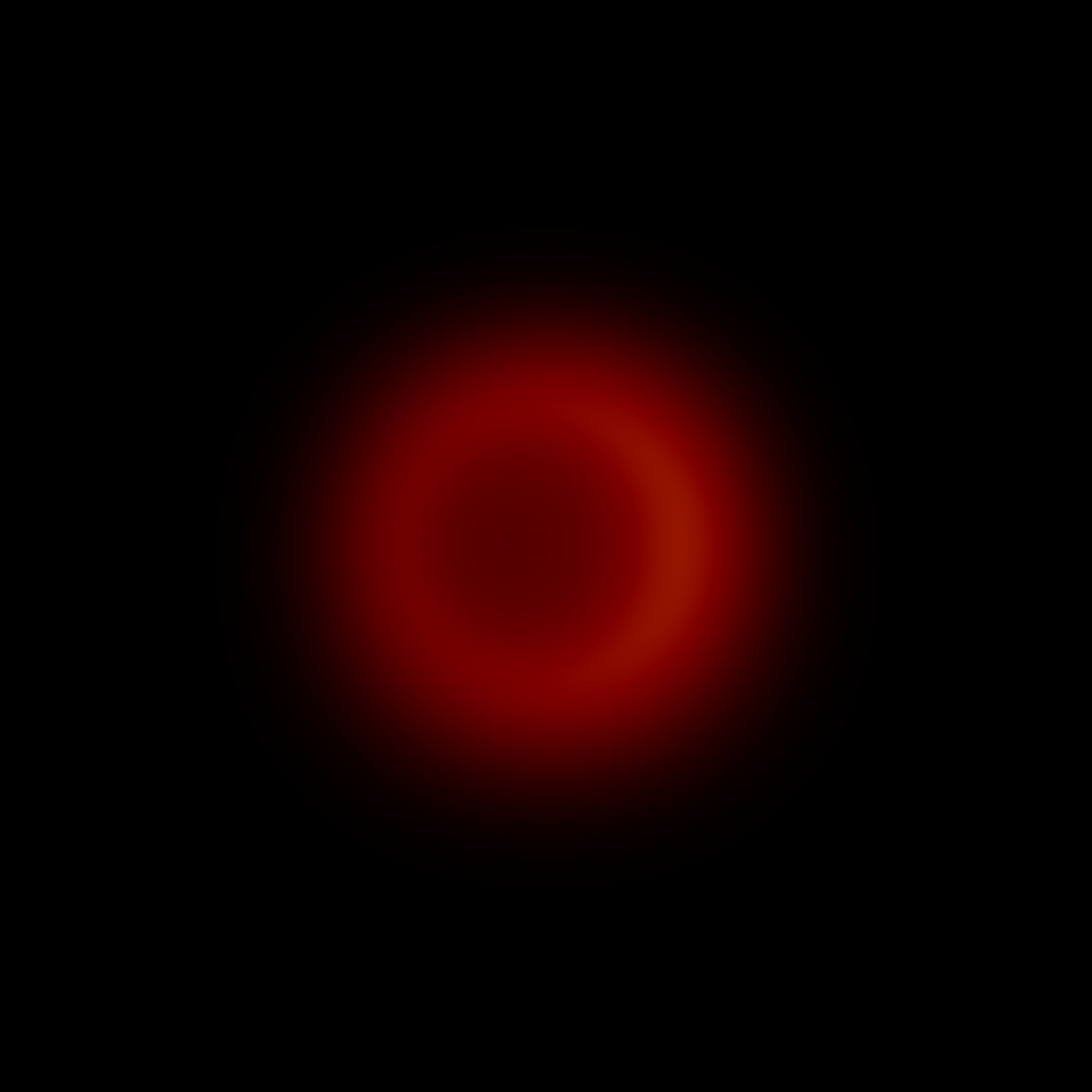}}
\\
\subfloat[EFLS5]{
\includegraphics[scale=0.45]{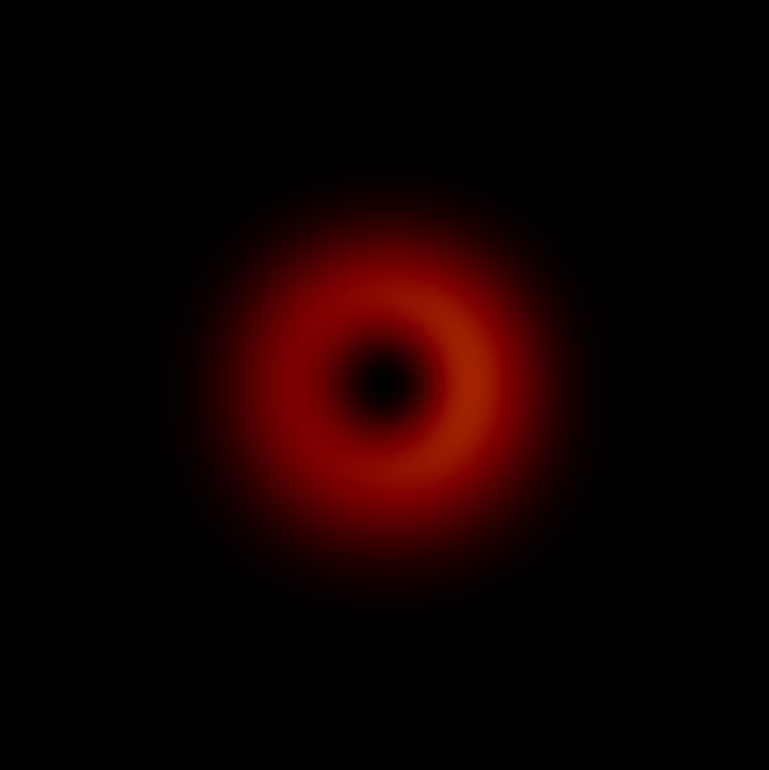}}
\subfloat[EFLS6]{
\includegraphics[scale=0.45]{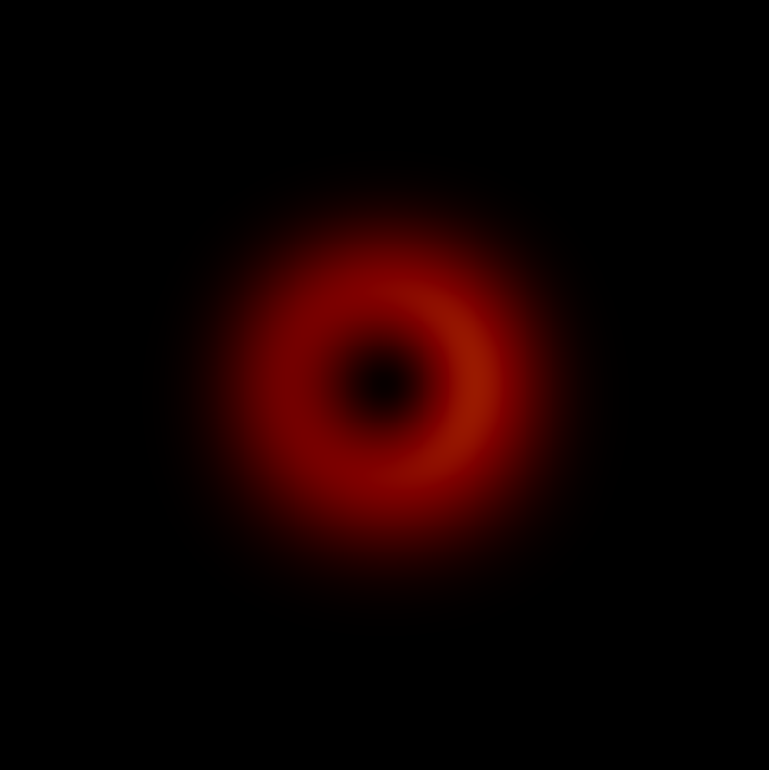}}

\caption{Intensity map of some E-FLS stars surrounded by an optically thick accretion disk, with no inner edge, extending until the origin. The observer location for this simulation is  $r_{obs} = 20 M $ and $\theta_{\textrm{obs}}=5^\circ$.}
    \label{newdisk_Lensing_EFLS2}
\end{figure*}

\begin{figure*}
\subfloat[$\theta_{\textrm{obs}}=5^\circ$]{
\includegraphics[scale=0.30]{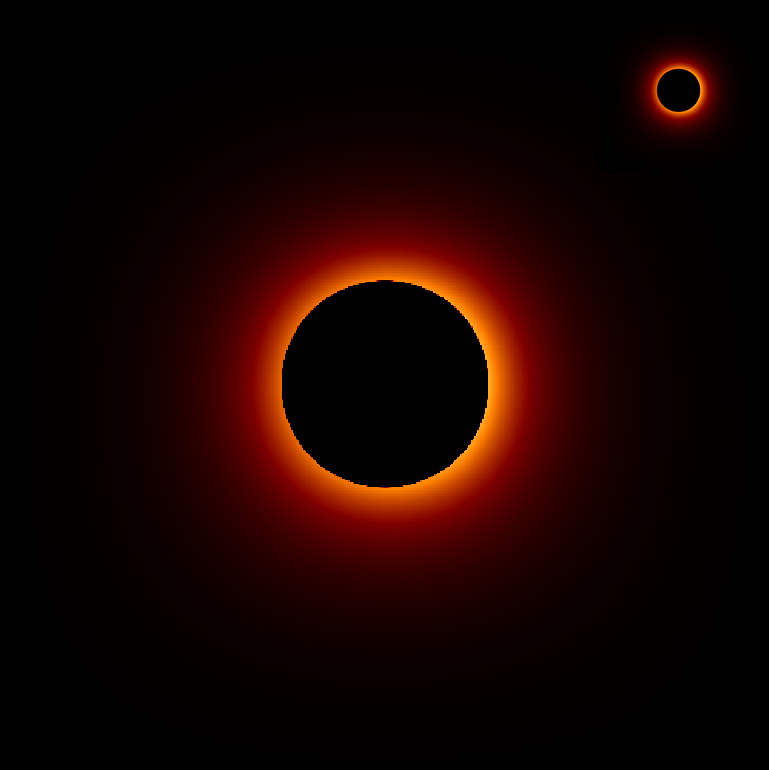}}
\subfloat[$\theta_{\textrm{obs}}=80^\circ$]{
\includegraphics[scale=0.30]{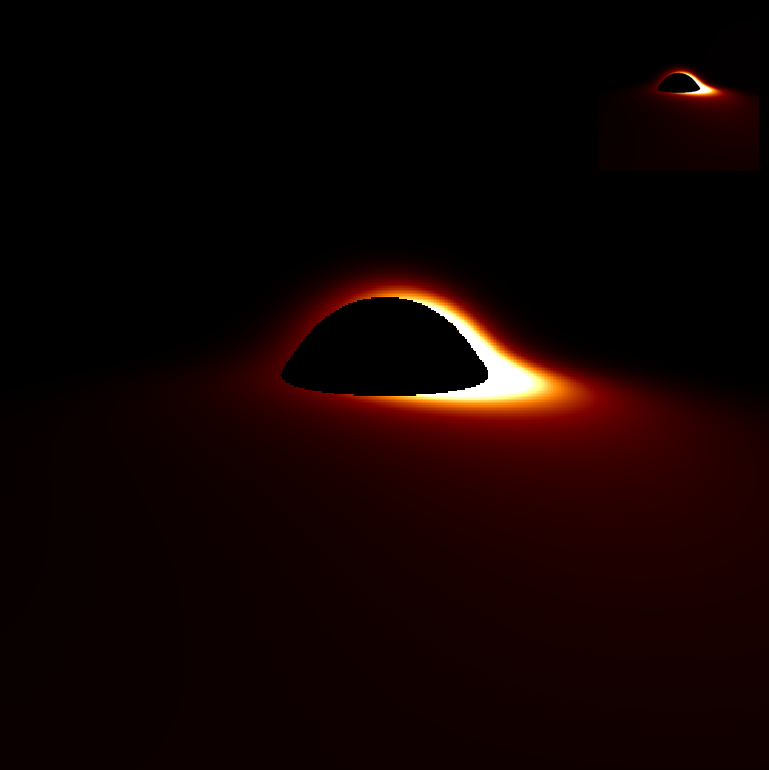}}

\caption{Left panel: Intensity maps of configurations EFLS5 and EFLS6, shown on comparable scales, for the observational angle $\theta_{\textrm{obs}}=5^\circ$. Right Panel: Intensity maps of configurations EFLS5 and EFLS6, shown on comparable scales, for the observational angle $\theta_{\textrm{obs}}=80^\circ$. Both solutions present the same value of $\mu$ and $\phi_{0}$, given by $\mu = 0$ and $\phi_{0}=1.15.$}
\label{scaled_images_thick}
\end{figure*}

\subsection{Astrophysical images of self-interacting E-FLS stars surrounded by an optically thick disk}
We first examine the case of a geometrically thin, optically thick accretion disk located at the equatorial plane of the E-FLS star. Due to the disk's opacity, light rays are evolved from the observer until they either escape to infinity or hit the accretion disk. The disk is modeled as consisting of particles in circular orbits, with the energy, angular momentum, and orbital frequency at a given radius $r_c$ provided in Eqs.~\eqref{E-L-timelike} and \eqref{Omega-timelike}. We model the radiation as monochromatic, with emission frequency $\nu_{em}$:
\begin{align}
\label{em_profile}I_\nu\propto \delta(\nu-\nu_{em})\,\epsilon(r),
\end{align}
where $ \delta(\nu-\nu_{em})$ is the Dirac delta function and $\epsilon(r)$ is the radial emission profile. We choose to work with the emission profile defined by~\cite{Proca_Star_Shadow, Vincent2022}:
\begin{align}
\label{em_profile2}\epsilon_{isco}(r)\equiv \frac{1+\tanh[50(r-6M)]}{2}\left(\frac{6\,M}{r}\right)^3.
\end{align}
This profile ensures a steep decay at large radii, consistent with standard thin-disk models, and introduces an inner cut-off radius. The cut-off in the emission profile at 
$r=6M$ is imposed to mimic that of a Schwarzschild black hole, which possesses an innermost stable circular orbit (ISCO) at this radius. Our goal is to assess whether a given emission profile around a BS can reproduce a black hole–like shadow. 

Since massive particles can circularly orbit the star very close to the center, we also adopt another emission profile that extends all the way to the origin, given by~\cite{Proca_Star_Shadow}:
\begin{align}
\label{em_profile3}\epsilon_{c}(r)\equiv e^{-r^{2}/50M^{2}}.
\end{align} 

Using the invariant quantities in Eq. \eqref{invar_quat}, we can compute how the observer measures the specific intensity, using
\begin{align}
\label{Iobs-Iem}I^{obs}_{\nu'}=\frac{\nu'^3}{\nu^3}I_{\nu},
\end{align}
where $\nu'$ is the frequency observed. In Fig.~\ref{Lensing_EFLS}, we show the astrophysical images for four E-FLS stars obtained with the emission profile~\eqref{em_profile2}, as seen by an observer placed at $r_{obs} = 20\,M$. In Fig.~\ref{Lensing_EFLS}, the images of EFLS3 and EFLS4 exhibit the weakest lensing effects, as these configurations do not possess closed circular photon orbits. Nevertheless, their gravitational fields are still strong enough to bend light, allowing the observer to discern the edges of the accretion disk around the star. For EFLS5 and EFLS6, the stars do support circular photon orbits; however, these features are not directly visible in the images due to the optical thickness of the disk. In all cases, the rotational motion of the disk induces a Doppler effect, producing a brightness asymmetry that makes the side of the disk moving toward the observer appear more luminous. Figure~\ref{Lensing_EFLS2} presents the images of the same configurations as seen from an inclination angle of $\theta = 5^\circ$, highlighting how the observer’s viewing angle modifies the apparent shape and contrast of the shadows. 

In Fig.~\ref{newdisk_Lensing_EFLS}, we present the astrophysical images considering the disk modeled with the emission profile given by Eq.\eqref{em_profile3}. We notice that, even though the disk extends from the origin to far away, the solutions with stronger lensing effects exhibit a dark region at the center. This behavior is even more pronounced for the images observed from the angle $\theta_{\textrm{obs}} = 5^\circ$, as shown in Fig.\ref{newdisk_Lensing_EFLS2}.





If we consider the characteristic scales defined by each star’s shadow, the configurations with self-interaction term $\lambda = 300$ are significantly larger than those with $\lambda = 0$. To compare their sizes, we construct an inset proportional to the scales of both configurations. Each shadow image was first computed in dimensionless coordinates, normalized by its own ADM mass. We then rescaled the EFLS5 shadow into the coordinate system of EFLS6 using the mass ratio,
\begin{equation}
\bar{x}= \frac{M_{\text{EFLS5}}}{M_{\text{EFLS6}}}x, \quad \bar{y} = \frac{M_{\text{EFLS5}}}{M_{\text{EFLS6}}}y,
\end{equation}
where $\bar{x}$ and $\bar{y}$ are the rescaled coordinates of EFLS5 in the coordinate system of EFLS6. This scaling shows that the shadow of EFLS5 occupies only $20.97\%$ of the size of EFLS6. The difference in the size of the astrophysical images is displayed in Fig.~\ref{scaled_images_thick}, which shows both shadows plotted with consistent scaling. The figure confirms that adding the self-interaction term yields configurations with substantially larger scales, consistent with the enhanced compactness and mass supported by the self-interacting case.

\subsection{Astrophysical Images of self-interacting E-FLS stars surrounded by an optically thin disk}

We now consider the astrophysical images of E-FLS stars surrounded by optically thin accretion disks. In this scenario, the disk is transparent to radiation, allowing light rays to cross it multiple times, with each crossing contributing to the increasing of the observed intensity. We adopt the same emission profile as in Eqs.~\eqref{em_profile} - \eqref{em_profile2}. The observer is placed at $r_{obs} = 20\,M$, while the polar angle $\theta_{obs}$ is varied. 

\begin{figure*}
\subfloat[EFLS3]{
\includegraphics[scale=0.45]{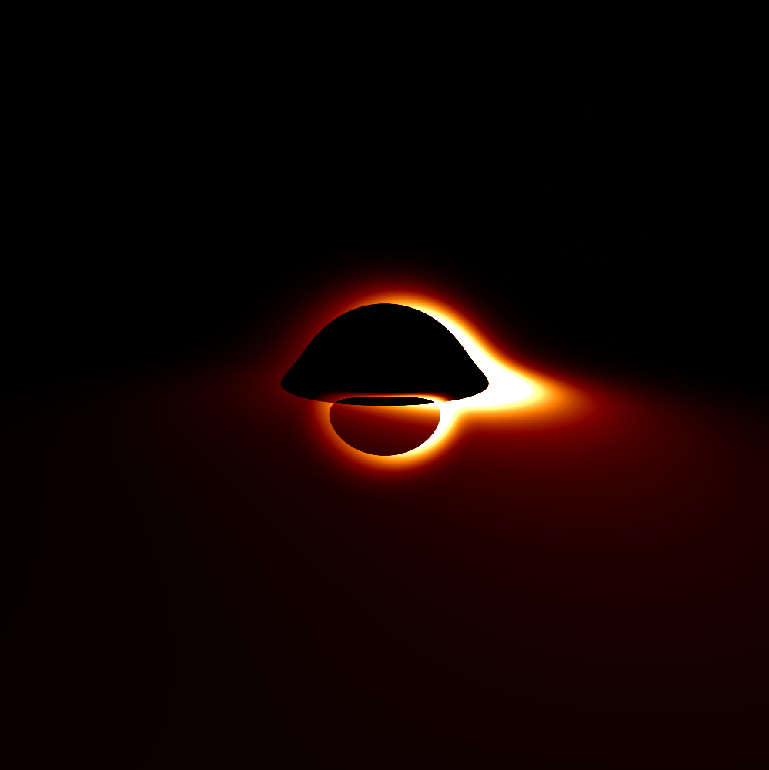}}
\subfloat[EFLS4]{
\includegraphics[scale=0.45]{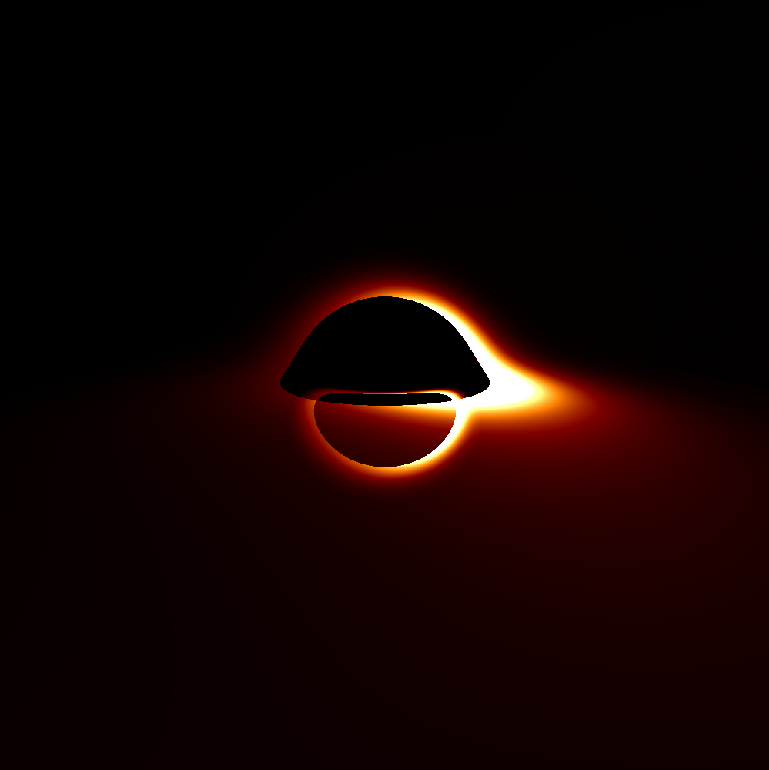}}
\\
\subfloat[EFLS5]{
\includegraphics[scale=0.45]{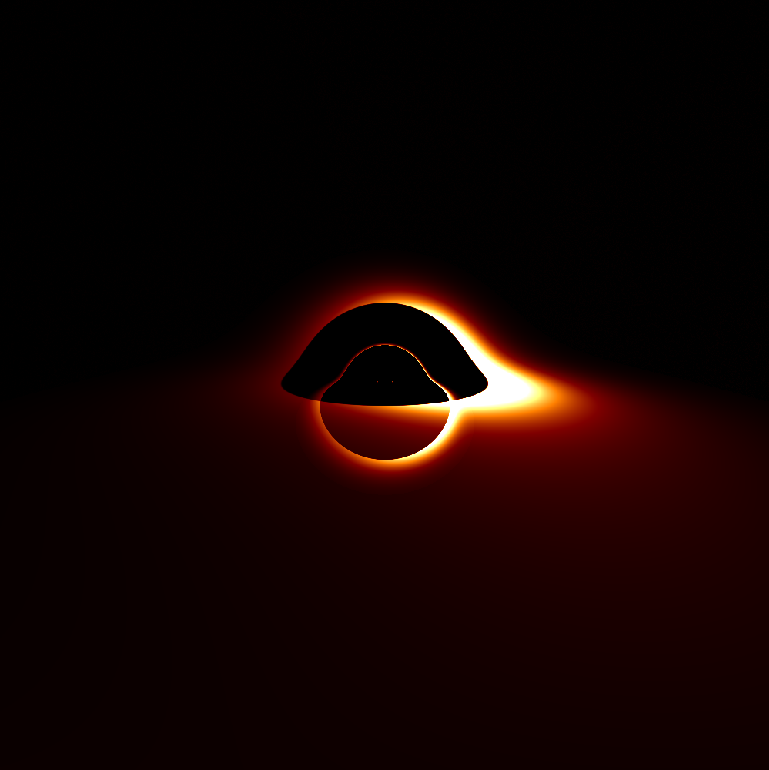}}
\subfloat[EFLS6]{
\includegraphics[scale=0.45]{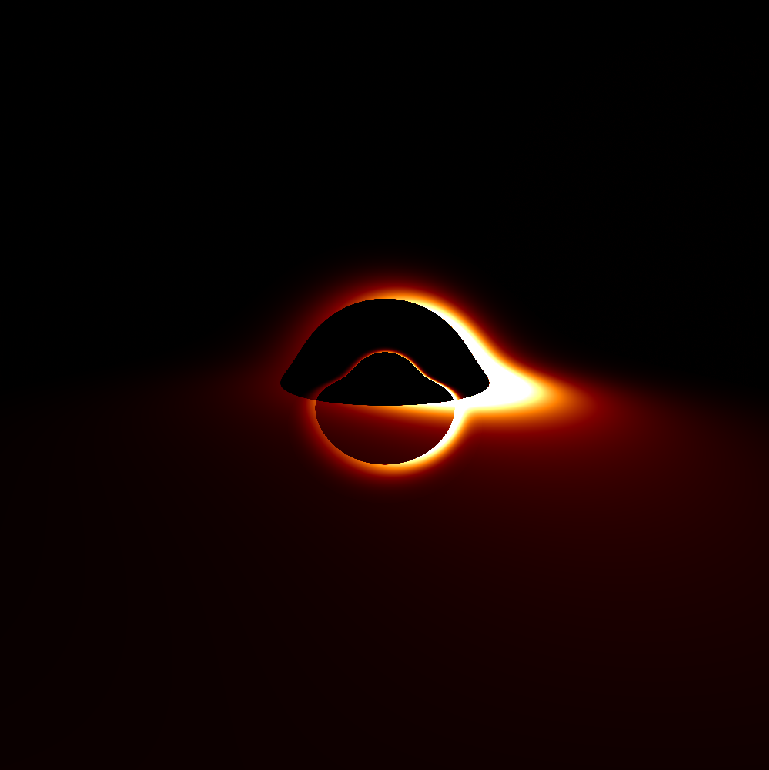}}

\caption{Intensity map of some E-FLS stars surrounded by an optically thin accretion disk. The observer location in the simulation is set to $r_{obs} = 20 M $ and $\theta_{\textrm{obs}}=80^\circ$. The optically thin nature of the accretion disk allows the astrophysical images to present distinct characteristic for each solution.}
    \label{Lensing_thin_EFLS}
\end{figure*}

In Fig.~\ref{Lensing_thin_EFLS}, we show the astrophysical images for an observer at $\theta_{obs} = 80^\circ$. The central regions of the images reveal features related to specific properties of the chosen stars. Configurations EFLS3 and EFLS4 exhibit qualitatively similar shadows. However distinct intensity distributions can be noticed when comparing their carefully, specially close to the center of the image. For EFLS5 and EFLS6, configurations which possess light rings, the intensity close to the center of the image is notably different. The presence of a photon ring induces a bright, nearly circular structure at the image center.

Figure~\ref{Lensing_EFLS2} presents images for an observer at $\theta_{obs} = 5^\circ$. In these images, concentric bright rings are visible near the center, with the largest corresponding to the primary photon ring. Additional inner rings emerge due to the absence of an event horizon, which would otherwise prevent photons from escaping after multiple scatterings. The distribution and multiplicity of these rings depend sensitively on the model parameters, providing a potential diagnostic tool for distinguishing among E-FLS star configurations.


     
\begin{figure*}
\subfloat[EFLS3]{
\includegraphics[scale=0.45]{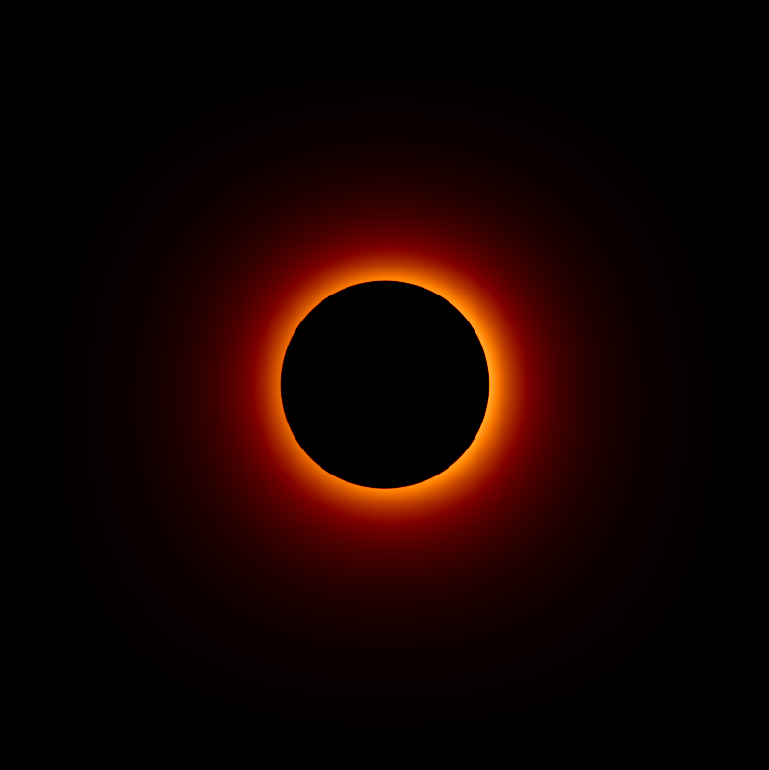}}
\subfloat[EFLS4]{
\includegraphics[scale=0.45]{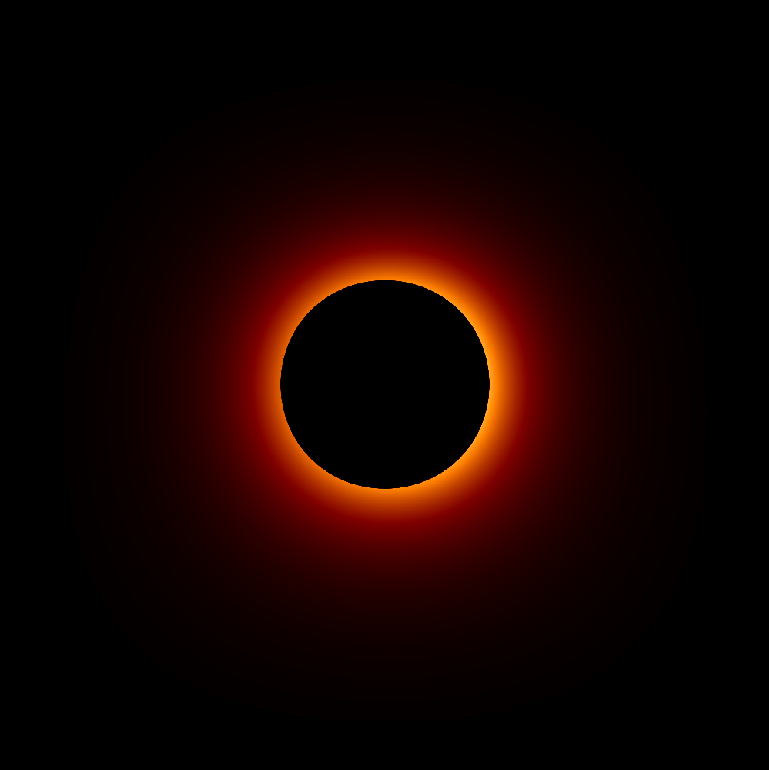}}
\\
\subfloat[EFLS5]{
\includegraphics[scale=0.45]{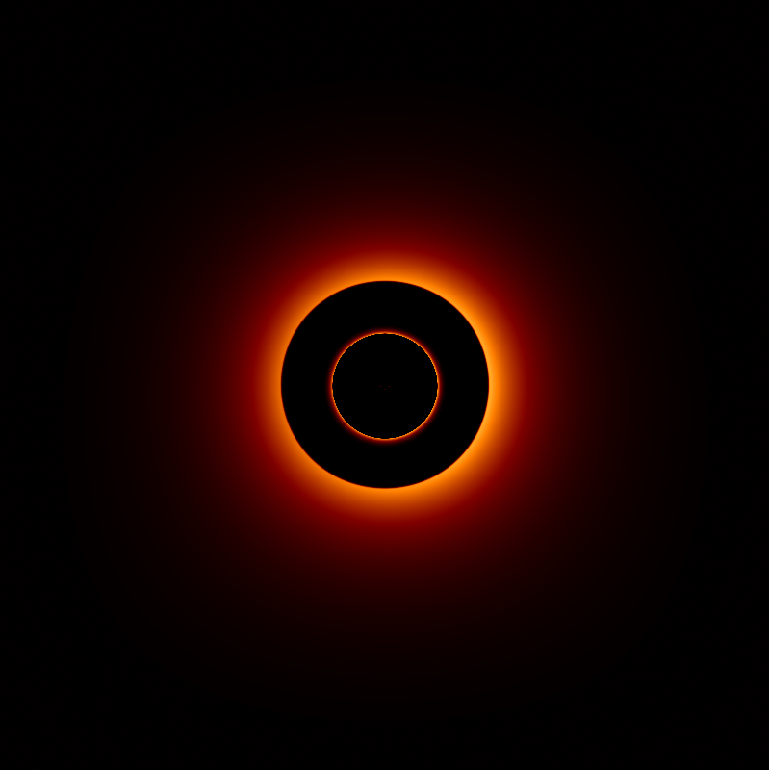}}
\subfloat[EFLS6]{
\includegraphics[scale=0.45]{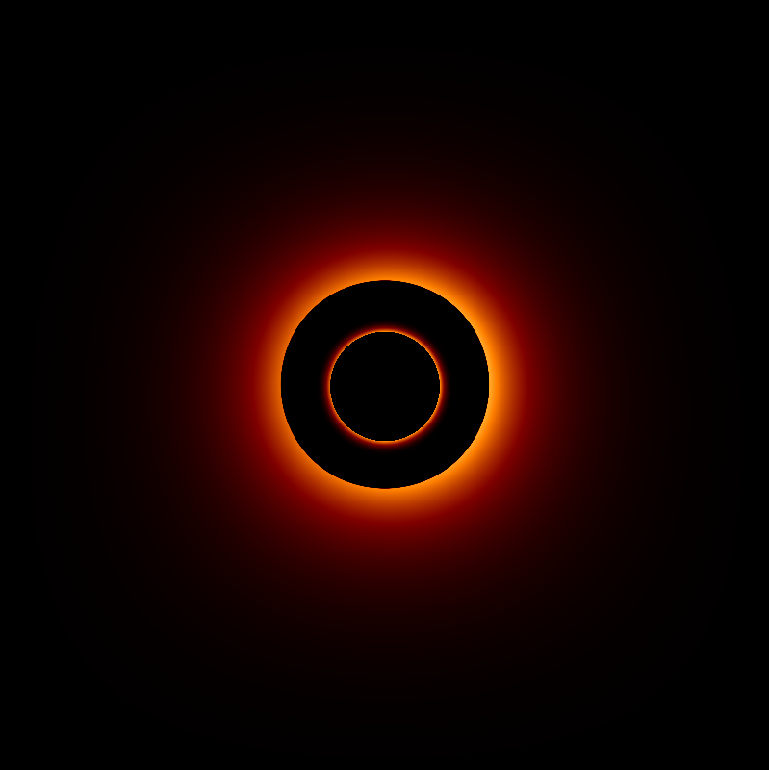}}

\caption{Intensity map of some E-FLS stars surrounded by an optically thin accretion disk. The observer location is set to $r_{obs} = 20 M $ and $\theta_{\textrm{obs}}=5^\circ$. The optically thin accretion disk allow the existence of photon rings near the center of the image.}
    \label{Lensing_EFLS2_2}
\end{figure*}


\begin{figure*}
\subfloat[$\theta_{\textrm{obs}}=5^\circ$]{
\includegraphics[scale=0.30]{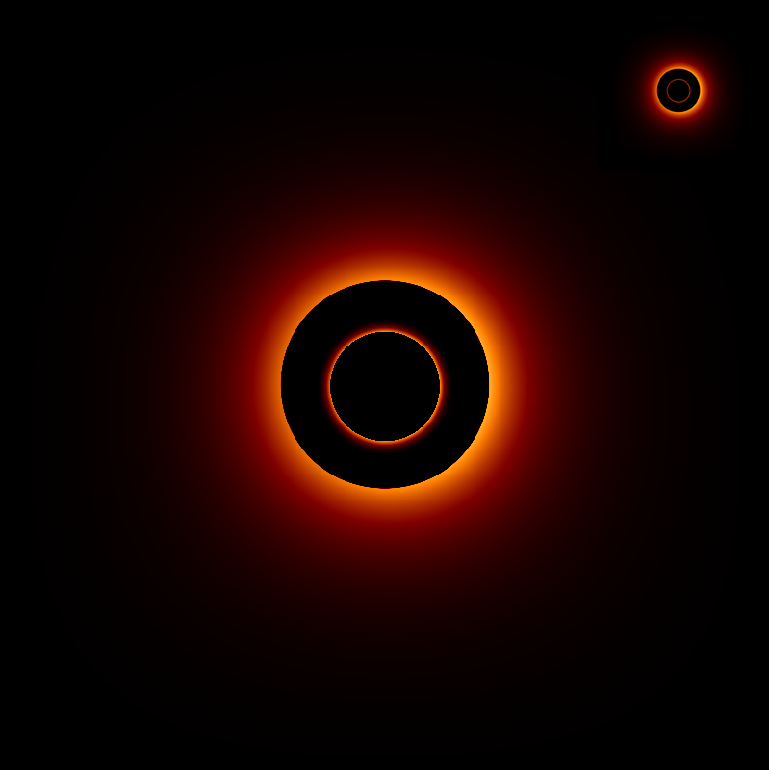}}
\subfloat[$\theta_{\textrm{obs}}=80^\circ$]{
\includegraphics[scale=0.30]{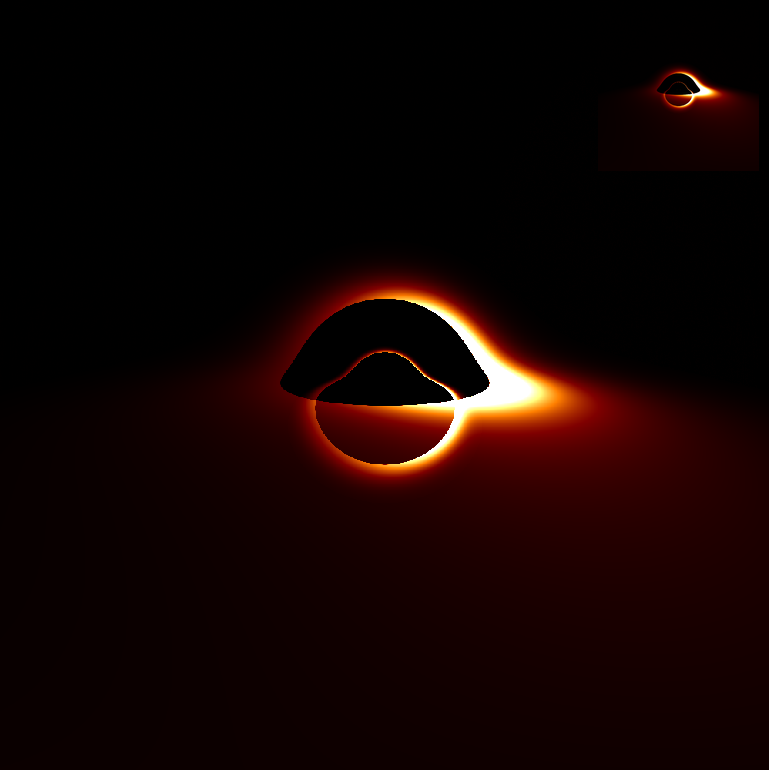}}

\caption{Left panel: Intensity maps of configurations EFLS5 and EFLS6, shown on comparable scales, for the observational angle $\theta_{\textrm{obs}}=5^\circ$. Right Panel: Intensity maps of configurations EFLS5 and EFLS6, shown on comparable scales, for the observational angle $\theta_{\textrm{obs}}=80^\circ$. Both solutions present the same value of $\mu$ and $\phi_{0}$, given by $\mu = 0$ and $\phi_{0}=1.15.$}
    \label{scaled_images_thin}
\end{figure*}

\section{Conclusions}
\label{sec:conclusions}

In this work, we investigated static self-interacting boson star solutions within the E-FLS model, dubbed as self-interacting E-FLS stars. We employed numerical techniques to solve the E-FLS field equations, assuming spherical symmetry for both the metric and scalar fields, enabling a complete exploration of the solution space. We analyzed the E-FLS star solutions by varying the central value of the complex scalar field $\phi_{0}$, the mass parameter $\mu$, and the self-interaction term $\lambda$. The variation of these parameters allowed us to probe the behavior of the solutions across different regimes, from weakly to strongly self-interacting configurations, and to assess the influence of each parameter on the resulting boson star properties.

The solutions exist within a bounded frequency range that shifts as $\lambda$ increases. The ADM mass exhibits two distinct behaviors depending on $\mu$: for $\mu = 0$, the solution curve remains isolated, whereas for $\mu \neq 0$ the solutions tend to merge as $\lambda$ grows. In all cases, the maximum ADM mass increases with $\lambda$. Stability analysis based on the binding energy $E_{b}$ indicates that configurations with negative bound energy increases for $\mu = 0$ and larger values of $\lambda$. Furthermore, both the effective radius $R_{\mathrm{eff}}$ and the compactness show dependence on the parameters $\mu$ and $\lambda$.

We analyzed null geodesics around selected E-FLS star configurations, choosing the solutions EFLS3, EFLS4, EFLS5 and EFLS6 in Table I. The existence of circular photon orbits depends on the value of the complex scalar field at the origin. For larger values of the complex scalar field at the origin, the solutions are more compact and therefore may admit circular photon orbits. Configurations EFLS1 through EFLS4 in Table I, although corresponding to the maximum mass solutions for fixed $\mu$ and $\lambda$, are not compact enough to support circular photon orbits. In contrast, configurations EFLS5 and EFLS6 are sufficiently compact to allow for the existence of circular photon orbits. This presence/absence of circular photon orbits significantly impacts the bending of light around the stars. We computed the null geodesics around these stars and found that configurations EFLS5 and EFLS6 strongly deflect light rays and support circular photon orbits, in contrast to the other solutions.  

To investigate the astrophysical images of these objects, when surrounded by an accretion disk, we applied the backwards ray-tracing method. For observers slightly displaced from the equatorial plane, the images clearly show the influence of gravitational lensing due to the E-FLS stars. The gravitational lensing effect is weak for configurations like EFLS3 and EFLS4. However, for configurations that are sufficiently compact to support circular photon orbits, such as EFLS5 and EFLS6, strong gravitational lensing effects are observed. The Doppler effect caused by the rotation of the accretion disk results in brightness asymmetry in the observed images. The presence of a significant self-interaction term can also lead to configurations with a much larger mass and effective radius. In order to make clear how these stars can differ in size, we also compared the shadows of the configurations EFLS5 and EFLS6 considering their proportional dimensions, finding that configurations with a larger self-interaction term produce larger shadows.

\begin{acknowledgments}
The authors would like to acknowledge Funda\c{c}\~ao Amaz\^onia de Amparo a Estudos e Pesquisas (FAPESPA), Conselho Nacional de Desenvolvimento Cient\'ifico e Tecnol\'ogico (CNPq) and Coordena\c{c}\~ao de Aperfei\c{c}oamento de Pessoal de N\'ivel Superior (CAPES) -- Finance Code 001, from Brazil, for partial financial support. Additionally, L. C. would like to thank the University of Aveiro, in Portugal, for the kind hospitality during the completion of this work. This work is supported by CIDMA under the Portuguese Foundation for Science and Technology (FCT, \url{https://ror.org/00snfqn58}) Multi-Annual Financing Program for R\&D Units, grants UID/4106/2025,  UID/PRR/4106/2025”, 2022.04560.PTDC (https://doi.org/10.54499/2022.04560.PTDC) and 2024.05617.CERN (https://doi.org/10.54499/2024.05617.CERN). 
This work has further been supported by the European Horizon Europe staff exchange (SE)
programme HORIZON-MSCA-2021-SE-01 Grant No. NewFunFiCO-101086251. 
\end{acknowledgments}
{}
\end{document}